\documentclass[11pt]{article}

\usepackage[preprint]{acl}

 \usepackage{booktabs}
\usepackage{threeparttable}
\usepackage{amsmath}
\usepackage{amssymb}
\usepackage{twemojis}
\usepackage{graphicx}
\usepackage[table,dvipsnames]{xcolor}
\usepackage{multirow}
\usepackage{booktabs}
\usepackage{ragged2e}
\usepackage{pmboxdraw}
\usepackage{tcolorbox}
\usepackage{stfloats}
\usepackage{float}
\usepackage{caption}
\usepackage{cuted}
\usepackage{placeins}
\usepackage{multicol}
\usepackage{enumitem}
\usepackage{relsize}

\newcommand{\midnotesize}{%
  \fontsize{7.5pt}{9pt}\selectfont
}

\makeatletter

\newcommand{\sub}{\quad\textcolor{gray}{\textSFii}\;}

\usepackage[english,bidi=default]{babel} % English as the main language.
\babelfont{rm}{TeXGyreTermesX} % similar to Times
\babelprovide[import]{hindi}
\babelfont[*devanagari]{rm}{Lohit Devanagari}
\babelprovide[import]{arabic}
\babelfont[*arabic]{rm}{Noto Sans Arabic}

\newcommand{\cat}[2]{#1\,\twemoji{#2}}

\usepackage{tabularx}
\usepackage{arydshln}
\usepackage{threeparttable}

\usepackage{listings}
\usepackage[capitalise]{cleveref}
\crefname{figure}{Figure}{Figures}
\crefname{table}{Table}{Tables}
\crefname{appendix}{Appendix}{Appendices}

\usepackage{mdframed}
\usepackage{listings}
\usepackage{xcolor}

\definecolor{lstbg}{HTML}{ffffff}
\definecolor{lstframe}{HTML}{D6D6D6}
\definecolor{lsttitle}{HTML}{37474F}
\definecolor{lstaccent}{HTML}{1565C0}

\lstdefinestyle{promptcompact}{
  basicstyle=\ttfamily\tiny\linespread{0.96}\selectfont,
  columns=fullflexible,
  breaklines=true,
  breakindent=0pt,
  breakatwhitespace=true,
  postbreak=\mbox{\textcolor{lsttitle}{\(\hookrightarrow\)}\space},
  keepspaces=true,
  showstringspaces=false,
  frame=none,
  aboveskip=0pt,
  belowskip=0pt,
}

\newmdenv[
  backgroundcolor=lstbg,
  roundcorner=4pt,
  innerleftmargin=5pt,
  innerrightmargin=4pt,
  innertopmargin=3pt,
  innerbottommargin=3pt,
  skipabove=2pt,
  skipbelow=2pt,
  leftline=true,
  topline=false,
  bottomline=false,
  rightline=false,
  linecolor=lstaccent,
  linewidth=1.2pt,
]{promptboxC}

\newcommand{\prompttitleC}[1]{%
  \vspace{1pt}\noindent
  {\scriptsize
    \setlength{\fboxsep}{2pt}%
    \colorbox{lstaccent!12}{\textcolor{lstaccent!85!black}{\textbf{#1}}}%
  }\par\vspace{1pt}%
}

\newmdenv[
  backgroundcolor=white,
  linecolor=lstframe,
  linewidth=0.35pt,
  roundcorner=5pt,
  innerleftmargin=6pt,
  innerrightmargin=6pt,
  innertopmargin=6pt,
  innerbottommargin=4pt,
  skipabove=2pt,
  skipbelow=2pt,
]{promptpanel}

\newcommand{\promptsep}{\vspace{3pt}\hrule\vspace{5pt}}

\usepackage{todonotes}
\usepackage{soul}

\definecolor{TodoColor}{rgb}{1,0.7,0.6}

\def\adl@drawiv#1#2#3{%
        \hskip.5\tabcolsep
        \xleaders#3{#2.5\@tempdimb #1{1}#2.5\@tempdimb}%
                #2\z@ plus1fil minus1fil\relax
        \hskip.5\tabcolsep}
\newcommand{\cdashlinelr}[1]{%
  \noalign{\vskip\aboverulesep
           \global\let\@dashdrawstore\adl@draw
           \global\let\adl@draw\adl@drawiv}
  \cdashline{#1}
  \noalign{\global\let\adl@draw\@dashdrawstore
           \vskip\belowrulesep}}
\makeatother

\usepackage{xcolor}
\usepackage{xspace}

\newcommand{\badge}[2]{%
  \begingroup
  \setlength{\fboxsep}{1pt}%
  \colorbox{#1!18}{\textcolor{#1!85!black}{\scalebox{0.95}{\textsf{#2}}}}%
  \endgroup\xspace
}

\newcommand{\badgeblack}[2]{%
  \begingroup
  \setlength{\fboxsep}{1pt}%
  \colorbox{#1}{\textcolor{black}{\scalebox{0.95}{\textsf{#2}}}}%
  \endgroup\xspace
}

\newcommand{\Rtag}{\badge{blue}{Ref}}

\newcommand{\HTtag}{\badge{orange}{ASR\textsubscript{\texttt{T}}}}
\newcommand{\HLtag}{\badge{orange}{ASR\textsubscript{\texttt{L}}}}

\newcommand{\RHTtag}{\badge{teal}{Ref+ASR\textsubscript{\texttt{T}}}}
\newcommand{\RHLtag}{\badge{teal}{Ref+ASR\textsubscript{\texttt{L}}}}
\newcommand{\RHTLtag}{\badge{teal}{Ref+ASR\textsubscript{\texttt{T+L}}}}

\newcommand{\sectionrow}[1]{%
  \addlinespace[2pt]
  \multicolumn{10}{l}{\textit{#1}}\\[-2pt]
  \addlinespace[2pt]
}

\definecolor{bestcol}{HTML}{1565C0}
\definecolor{secondcol}{HTML}{B8860B}

\newcommand{\badgec}[2]{%
  \begingroup
  \setlength{\fboxsep}{1pt}%
  \colorbox{#1!18}{{\scalebox{0.95}{{\textbf{#2}}}}}%
  \endgroup\xspace
}

\newcommand{\best}[1]{\badgec{bestcol}{#1}}
\newcommand{\second}[1]{\badgec{secondcol}{#1}}

\definecolor{defcol}{HTML}{546E7A}   % blue-gray
\definecolor{iclcol}{HTML}{2E7D32}   % green
\definecolor{chunkcolstrat}{HTML}{6A1B9A} % purple
\definecolor{casccol}{HTML}{1565C0}  % blue
\definecolor{tcol}{HTML}{5E35B1}     % indigo
\definecolor{tacol}{HTML}{00897B}    % teal
\definecolor{loracol}{HTML}{D84315} % orange-reddish (deep orange)
\definecolor{fullcol}{HTML}{C2185B}

\definecolor{chunkcol}{HTML}{f7e1df}
\definecolor{audioencol}{HTML}{cee6ff}
\definecolor{windowgroupcol}{HTML}{e4cfec}
\definecolor{localsegmenttransformcol}{HTML}{bee5dd}
\definecolor{documentenccol}{HTML}{f2e8d8}

\newcommand{\DefaultStrat}{\badge{defcol}{Default}}
\newcommand{\ICLStrat}{\badge{iclcol}{ICL}}
\newcommand{\ChunkStrat}{\badge{chunkcolstrat}{Chunk}}
\newcommand{\Tonly}{\badge{tcol}{T}}
\newcommand{\TAudio}{\badge{tacol}{T{+}A}}
\newcommand{\SelfCasc}[1]{\badge{casccol}{Self-casc.}\,#1}
\newcommand{\LoRAStrat}{\badge{loracol}{LoRA}}

\definecolor{taskvarcol}{HTML}{C62828}

\newcommand{\FullStrat}{\badge{fullcol}{Fully FT}}

\newcommand{\NoTranscript}{\badge{taskvarcol}{-Tr.}}
\newcommand{\WTranscript}{\badge{taskvarcol}{+Tr.}}

\newcommand{\huggingfacesmall}{\includegraphics[width=9pt]{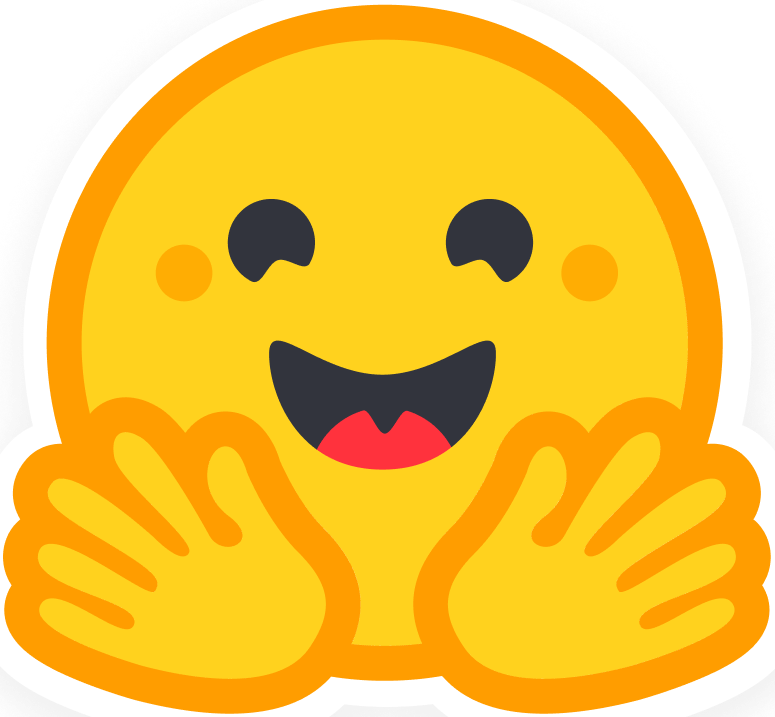}}

\usepackage{colortbl}

\usepackage{footmisc} % to refer to footnote

\usepackage[protrusion=true]{microtype}

\title{Beyond Transcripts: A Renewed Perspective on Audio Chaptering}

\newcommand{\authorsep}{\quad}

\author{%
\textbf{Fabian Retkowski}$^1$\authorsep
\textbf{Maike Züfle}$^1$\authorsep
\textbf{Thai Binh Nguyen$^1$}\\
\textbf{Jan Niehues}$^1$\authorsep
\textbf{Alexander Waibel}$^{1,2}$\\
\textsuperscript{1}Karlsruhe Institute of Technology\authorsep
\textsuperscript{2}Carnegie Mellon University\\
\texttt{\{fabian.retkowski, maike.zuefle\}@kit.edu}\\
}

\begin{document}

\maketitle

\begin{abstract}
Audio chaptering, the task of segmenting long-form audio into coherent sections, is increasingly important for navigating podcasts, lectures, and videos. Despite its relevance, research remains limited and text-based, leaving key questions unresolved about leveraging audio information, handling ASR errors, and transcript-free evaluation. We address these gaps through three contributions: (1) a systematic comparison between text-based models with acoustic features, a novel audio-only architecture (AudioSeg) operating on learned audio representations, and multimodal LLMs; (2) empirical analysis of factors affecting performance, including transcript quality, acoustic features, duration, and speaker composition; and (3) formalized evaluation protocols contrasting transcript-dependent text-space protocols with transcript-invariant time-space protocols. Our experiments on YTSeg reveal that AudioSeg substantially outperforms text-based approaches, pauses provide the largest acoustic gains, and MLLMs remain limited by context length and weak instruction following, yet MLLMs are promising on shorter audio.\footnote{\label{ft:links}We release the \texttt{chunkseg} evaluation package (\href{https://github.com/retkowski/chunkseg}{GitHub}, 
\href{https://pypi.org/project/chunkseg/}{PyPI}) and the AudioSeg model 
(\href{https://huggingface.co/retkowski/audioseg}{HF}); additional annotations have been merged into the YTSeg dataset repository (\href{https://huggingface.co/datasets/retkowski/ytseg}{HF}).}
\end{abstract}
\section{Introduction}

As long-form audio and video content becomes increasingly common, such as podcasts, lectures and YouTube videos, users need better tools to navigate and locate information within recordings. In practice, people rarely consume these recordings linearly \cite{yurum2023,liao2023}. Instead, they skim, scrub the timeline, jump to relevant moments, and return to specific sections. This non-linear behavior makes chapter markers a key interface for browsing and re-finding content \cite{Ghazimatin2024}. The task of \textit{audio chaptering} addresses this by automatically segmenting audio into coherent sections. Beyond navigation, chapters serve as a backbone for downstream applications such as summarization and question answering \cite{sarthi-2024,retkowski-etal-2025-summarizing,koneru-etal-2026-boom}.

Despite growing relevance, research on audio chaptering remains limited and predominantly text-based: models typically operate on transcripts and inherit evaluation protocols from text segmentation \cite{retkowski-waibel-2024-text,freisinger25_interspeech}. This transcript-centric framing leaves several key limitations unresolved. First, the role of audio remains unclear: because most prior work treats chaptering as a purely textual problem, there is limited understanding of how audio can be leveraged for chaptering, whether through hand-crafted acoustic features or learned representations, and whether it improves performance. Second, text-segmentation evaluation protocols assume a fixed transcript. In practice, chaptering systems often rely on ASR outputs whose errors and segmentation differences change the underlying unit sequence, especially the sentence boundaries and the number of sentences. This changes the granularity (and difficulty) of the segmentation task, so standard text-based metrics computed on different transcripts are not directly comparable and can appear to improve simply because one transcript is coarser, rather than because the model segments better. Finally, chapter boundaries are intrinsically defined in continuous time, but many pipelines ``snap'' these timestamps to sentence boundaries. This realignment is inherently lossy: it can shift boundaries away from their true temporal positions and can systematically bias evaluation toward sentence segmentation artifacts.

This work aims to establish a methodological foundation for audio chaptering, addressing aforementioned gaps through three main contributions:
\begin{enumerate}
    \item We systematically evaluate three modeling paradigms for audio chaptering: text-based models with and without acoustic feature augmentation, a novel audio-only architecture (AudioSeg) that operates directly on learned speech representations, and lastly, we explore whether multimodal large language models (MLLMs) are capable of this task.
    \item Second, we provide empirical insight into factors affecting chaptering performance. We analyze the robustness of text-based models to ASR errors, quantify the contribution of different acoustic features, and examine how audio characteristics such as duration and speaker composition influence segmentation.
    \item Third, we systemize evaluation for audio chaptering: we formalize existing text-based protocols, analyze their limitations, and introduce time-based evaluation that enables fair comparison across text-based, audio-only, and multimodal models independent of the transcript.
\end{enumerate}

To support further work on audio chaptering, we release an evaluation toolkit and AudioSeg.\footref{ft:links}
\section{Evaluation Protocols}
% \TODO{small intro}

\subsection{Text-Based Segmentation}

Segmentation has traditionally been studied in the text space $\mathcal{X}$, where a document is modeled as a sequence of discrete units (typically sentences). The objective is to identify a boundary sequence $\mathbf{y} = (y_1, \dots, y_{N-1})$, where $y_i = 1$ denotes a boundary between units $s_i$ and $s_{i+1}$. Evaluation compares a predicted sequence $\hat{\mathbf{y}}$ against a reference $\mathbf{y}$ using segmentation metrics such as $P_k$ \cite{beeferman_statistical_1999} and Boundary Similarity (B; \citealt{fournier-2013-evaluating}), as well as classification metrics such as F1 score. We adopt this formalism for audio chaptering. However, a fundamental domain mismatch exists: chapters are defined as continuous \emph{timestamps} in the time domain $\mathcal{T}$, whereas text segmentation assumes \emph{discrete indices}. The protocols below differ in how they map between $\mathcal{T}$ and $\mathcal{X}$, and whether the metric operates in $\mathcal{X}$ or $\mathcal{T}$.

\subsection{Text-Space Protocols}

%Research on audio chaptering is still limited, and evaluation practices are not yet fully standardized.
In existing work, evaluation is typically defined in text space by projecting continuous-time chapter boundaries onto transcript units, either on reference transcripts (R1) or on ASR transcripts (H1). We distinguish these two settings and then introduce two hybrid variants (H2--H3) that map ASR-based predictions back to a canonical reference transcript.

\subsubsection{Evaluation on Ref. Transcripts (R1)}

Let the \emph{reference transcript} be the canonical text representation of the audio, segmented into sentences
$S_{\text{ref}} = (s_1, \dots, s_N)$. Ground-truth chapter boundaries are given as timestamps $\mathcal{T}_{\text{gold}} \subset \mathbb{R}^+$ on the audio. To evaluate in text space, we first align $S_{\text{ref}}$ to the audio signal via forced alignment (FA) or closed caption timestamps to obtain a start time $t_{\text{start}}(s_i)$ and end time $t_{\text{end}}(s_i)$ for each sentence. We define a projection function $\phi_{\text{ref}}: \mathcal{T} \rightarrow \{1, \dots, N-1\}$ that maps each timestamp in $\mathcal{T}_{\text{gold}}$ to the nearest sentence boundary in $S_{\text{ref}}$.

In protocol \textbf{R1 (Ref)}, a model predicts a boundary sequence over sentences in the reference transcript, and we compute text segmentation metrics on $S_{\text{ref}}$. This approach has two implications: (1) continuous-time boundaries are projected to sentence boundaries (a lossy discretization); (2) evaluation assumes access to the transcript. This corresponds to the protocols used in most prior work \cite{lai_automatic_2016,retkowski-waibel-2024-text}.

\subsubsection{Evaluation on ASR Transcripts (H1)}

In realistic deployments, systems operate on an \emph{ASR transcript} rather than the reference. We therefore define protocol \textbf{H1 (ASR)}, which mirrors R1 but replaces the reference transcript with the ASR transcript. We segment the ASR output into sentences $S_{\text{asr}} = (u_1, \dots, u_M)$, where usually $M \neq N$, and obtain timestamps for each sentence (from the ASR decoder or via FA). Each chapter timestamp in $\mathcal{T}_{\text{gold}}$ is then mapped to the nearest ASR sentence boundary in $S_{\text{asr}}$, yielding a gold boundary sequence over ASR sentences. This protocol is also reflected in prior work \cite{lai_integrating_2020,freisinger25_interspeech}.

H1 removes the need for a reference transcript and directly reflects operating conditions, but scores become dependent on the particular ASR system used: different models or decoding settings can yield different segmentations, changing metrics even when the underlying time boundaries are identical. H1 is therefore \emph{not} transcript-invariant.

\subsubsection{Alignment to Reference Text (H2/H3)}

To recover a canonical evaluation space while still supporting models that operate on ASR, protocols \textbf{H2} and \textbf{H3} project ASR-based predictions back onto the reference transcript $S_{\text{ref}}$. Both establish a monotonic mapping from ASR sentences $S_{\text{asr}}$ to reference sentences $S_{\text{ref}}$, but differ in how this mapping is derived: \textbf{H2} uses token-level alignment between ASR and reference transcripts, while \textbf{H3} uses maximal temporal overlap between ASR and reference sentences. Once the mapping is established, predicted boundaries between ASR sentences are transferred to $S_{\text{ref}}$ and evaluated using standard text segmentation metrics. Both protocols anchor scores in a single canonical transcript, improving comparability across systems. H2 leverages lexical correspondence but is sensitive to ASR errors; H3 relies solely on timestamps, reducing dependence on word-level accuracy. See \Cref{app:alignment} for details. % of the projection.

\subsection{Time-Space Protocols}

\subsubsection{Discrete-Time Evaluation (T1)}

All text-space protocols above assume evaluation over a sequence of text units. For audio chaptering, however, the primary object is the \emph{time axis}. Protocol \textbf{T1 (Time-based, discrete)} evaluates segmentation in a discretized time space without privileging any transcript. The audio duration $D$ is discretized into fixed-length frames (e.g., $\Delta t = 1\text{s}$), yielding $K = \lceil D/\Delta t \rceil$ chunks. Both gold and predicted boundaries are mapped to these chunks, producing binary sequences of length $K$. This yields a boundary sequence over time chunks, directly analogous to a boundary sequence over sentences. Crucially, text segmentation metrics such as \(P_k\) operate on sequences of binary decisions and therefore apply unchanged when we replace sentences by time chunks. T1 accommodates any model whose outputs can be expressed as time intervals: audio-only models predicting on time chunks, models that output timestamps, and even text-based models, whose sentence-level boundaries can be mapped back to time via ASR timestamps or FA. Once outputs are mapped to time chunks, the transcript no longer enters the metric, making evaluation \emph{transcript-invariant}.

\subsubsection{Continuous-Time Evaluation (T2)}

Finally, protocol \textbf{T2 (Time-based, continuous)} evaluates segmentation directly in continuous time. We represent the reference and predicted segmentations by ordered sets of boundary timestamps, $\mathcal{T}_{\text{gold}} = \{t_1, \dots, t_K\}$ and $\hat{\mathcal{T}}_{\text{pred}} = \{\hat{t}_1, \dots, \hat{t}_L\}$. Unlike T1, text segmentation metrics are not applicable in continuous space. For time-stamped events, collar-based evaluation (e.g., F1 under a fixed temporal tolerance around each boundary) is common in sound event detection and speaker diarization \citep{2016_mesaros,bredin17_interspeech}. Early work applies such tolerance-based boundary scoring to topic segmentation \citep{guinaudeau10_interspeech}, but to our knowledge, this perspective has not been made explicit or systematized. T2 is the most direct formulation of the audio chaptering task: it requires no transcript, alignment, or discretization and produces transcript-invariant scores.
\section{Approaches}

We investigate three paradigms for audio chaptering: (1) text-based;  operating on transcripts, optionally augmented with acoustic features; (2) audio-only; without transcription; and (3) MLLMs. For text-based approaches, we build on MiniSeg \cite{retkowski-waibel-2024-text}, a hierarchical transformer that serves as our baseline.

\subsection{Text-Based Baseline}

MiniSeg \cite{retkowski-waibel-2024-text} is a two-stage hierarchical model for text segmentation. A trainable sentence encoder (based on MiniLM; \citealt{wang2020}) produces fixed-dimensional embeddings for each sentence in the transcript. These embeddings are then processed by a RoFormer document encoder, which performs sequence labeling to predict binary boundary decisions. The model is trained with weighted binary cross-entropy to address class imbalance.

\subsection{Hand-Crafted Audio Features}

\begin{table}[h]
\centering
\midnotesize
\setlength{\tabcolsep}{3pt}
\renewcommand{\arraystretch}{1.15}
\begin{tabularx}{\linewidth}{@{} l >{\RaggedRight\arraybackslash}X @{}}
\toprule
\textbf{Category} & \textbf{Features} \\
\midrule
\cat{Pauses}{hourglass} &
  \ttfamily pause\_duration \\
\cat{Speaking Rate}{stopwatch} &
  \ttfamily wpm, z\_wpm \\
\cat{Pitch}{musical note} &
  \ttfamily mean\_f0, std\_f0, min\_f0, max\_f0, range\_f0, slope\_f0, voicing\_ratio, zF0\_mean, zF0\_slope \\
\cat{Loudness}{speaker high volume} &
  \ttfamily lkfs, z\_loudness \\
\cat{Speakers}{busts in silhouette} &
  \ttfamily same\_as\_prev, speaker\_change, turn\_id, pos\_in\_turn, turn\_len, dist\_prev\_same, num\_speakers\_so\_far \\
\bottomrule
\end{tabularx}

\caption{Audio features extracted for each sentence}\vspace{-0.2cm}
\label{tab:features}
\end{table}

While text-based models rely primarily on the semantic content of transcripts, we hypothesize that incorporating acoustic and conversational cues can improve segmentation performance. We therefore augment MiniSeg with hand-crafted audio features extracted from the speech signal; the corresponding features are presented in \Cref{tab:features} while the extraction methodology is detailed in \Cref{app:feature_extraction}.

\subsubsection{Feature Fusion}

We integrate features using concatenation-based fusion. For each sentence, we concatenate the sentence embedding from MiniSeg's sentence encoder with the feature vector, then apply linear projection: %$\mathbf{h}_i = \text{Linear}([\mathbf{e}_i \,||\, \mathbf{f}_i])$, 
\begin{equation*}
\mathbf{h}_i = \text{Linear}([\mathbf{e}_i \,||\, \mathbf{f}_i]),
\end{equation*}
where $\mathbf{e}_i \in \mathbb{R}^{d_{\text{emb}}}$ is the sentence embedding, $\mathbf{f}_i \in \mathbb{R}^{d_{\text{feat}}}$ the feature vector, and $\mathbf{h}_i \in \mathbb{R}^{d_{\text{emb}}}$ the fused representation passed to the document encoder.

\subsection{Audio-Only Model}
\label{sec:audioseg}

\begin{figure*}[ht]
    \centering
    \includegraphics[width=0.75\linewidth]{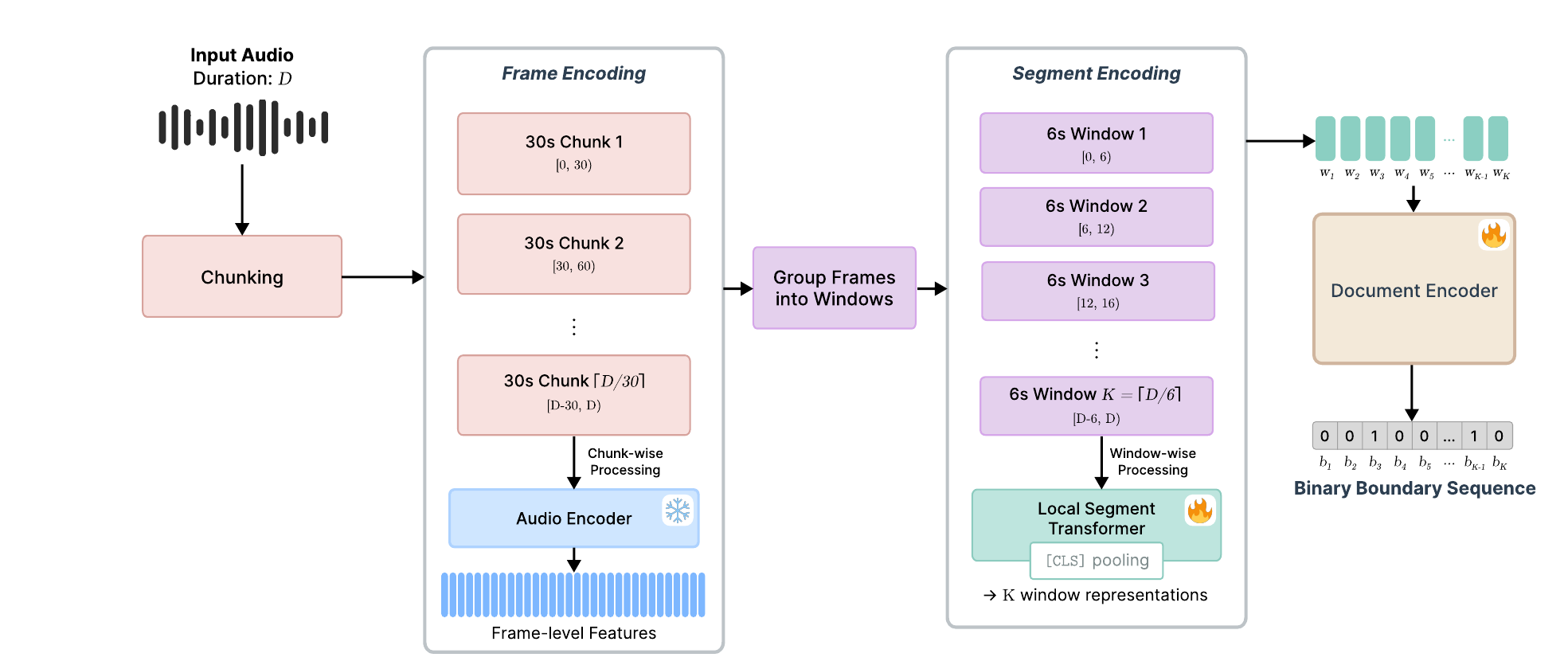}
\caption{\textbf{AudioSeg} processes input audio of duration $D$ through three stages: \textbf{Frame Encoding} extracts frame-level features from \badgeblack{chunkcol}{30s chunks} using a \badgeblack{audioencol}{frozen audio encoder} \twemoji{snowflake}; \textbf{Segment Encoding} groups frames into \badgeblack{windowgroupcol}{6s windows} and encodes each via a \badgeblack{localsegmenttransformcol}{Local Segment Transformer} \twemoji{fire} with [CLS] pooling to produce $K = \lceil D/\Delta t \rceil$ segment embeddings; \textbf{Document Encoding} processes the segment sequence through a \badgeblack{documentenccol}{RoFormer encoder} \twemoji{fire} to predict a binary boundary sequence $(b_1, \ldots, b_K)$ indicating chapter boundaries.}\vspace{-0.2cm}    \label{fig:audioseg}
\end{figure*}

We develop an audio-only segmentation model (\mbox{AudioSeg}) that operates directly on frame-level acoustic features without requiring a transcript. The architecture follows a three-stage processing pipeline (\Cref{fig:audioseg}): \textit{frame encoding} transforms raw audio into frame-level representations, \textit{segment encoding} aggregates these features into segment-level embeddings, and \textit{document encoding} models long-range dependencies to predict chapter boundaries.

\paragraph{Frame Encoding.} Audio is processed by a frozen, pre-trained audio encoder that produces frame-level features at a temporal stride. To handle long audio, we adopt chunked processing: the audio is divided into fixed-length chunks (e.g., 30~s), frame-level features are extracted independently for each chunk, and the resulting frames are concatenated to form a continuous sequence spanning the full audio.

\paragraph{Segment Encoding.} The frame sequence is partitioned into non-overlapping windows of duration $\Delta t$ (e.g., 6~s). Each window is encoded into a segment-level representation using a \textit{Local Segment Transformer} which projects frames to a hidden dimension, prepends a learnable \texttt{[SEG]} token, adds positional embeddings, and applies a local transformer encoder. The resulting \texttt{[SEG]} token is taken as the segment embedding, producing $K=\lceil D/\Delta t \rceil$ segment embeddings.

\paragraph{Document Encoding.} The segment embeddings are processed by a RoFormer encoder to model long-range dependencies across the full document. The outputs are passed through a linear layer to produce a boundary probability for each segment.

\paragraph{Training.} Ground-truth chapter boundaries (continuous timestamps) are discretized to the segment grid by assigning a boundary label to the segment containing each timestamp. The model is trained with binary cross-entropy loss on these labels. This formulation enables direct evaluation under protocol T1 (discrete-time) and straightforward mapping to T2 (continuous-time) by converting predicted segment boundaries back to timestamps.

\subsection{Multimodal LLMs}
Finally, we evaluate MLLMs to assess whether instruction-following, \textit{general-purpose} models can perform the audio chaptering task. Unlike the previously considered approaches, which focus solely on segmentation, MLLMs can be prompted to jointly perform multiple subtasks end-to-end in a single pass. Our main setting requires joint transcription, segmentation, and chapter title generation. We investigate the following prompting strategies and adaptations.

\paragraph{Zero-Shot.}
We evaluate the models in a zero-shot setting, prompting them to transcribe alongside chapter boundaries and titles using tags.

\paragraph{Chunking.}
To mitigate limitations for long audio, we experiment with splitting the audio into 30 second segments, and for each, the model is provided with the audio chunk along with the textual output from previous chunks, and prompted to continue the transcript with chapter annotations.

\paragraph{In-Context Learning (ICL).}
To further improve task understanding and output formatting, we evaluate in-context learning (ICL; \citealt{brown_fewshot_2020}).

\paragraph{Self-Cascaded.}
To disentangle whether the observed performance of MLLMs arises from their multimodal capabilities or from their underlying language modeling capacity, we evaluate two self-cascaded variants: (1) the model is first prompted to generate a transcript, and in a second step is asked to insert chapters based solely on the transcript; and (2) the model first generates a transcript, and in a second step is provided with both the audio and transcript to insert chapter annotations.

\paragraph{LoRA Training.}
Finally, we fine-tune the MLLM using LoRA \cite{hu2021loralowrankadaptationlarge} to evaluate if lightweight task-specific adaptation improves performance over prompting-based approaches.

\subsubsection{Temporal Grounding} In addition to the main setting, we explore two alternative task formulations to investigate whether chapters can be temporally grounded: (1)~\textbf{Timestamps + Titles}, where the model predicts chapter boundaries and titles directly from audio without generating a transcript. This is a particularly efficient formulation as it avoids generating all transcription tokens while still producing a segmentation; and (2)~\textbf{Transcription + Timestamps + Titles}, which extends the main setting by additionally requiring timestamps for each chapter boundary, combining transcription, segmentation, temporal grounding, and title generation in a single pass.

\section{Experimental Setup}
% In this section, we detail training and evaluation data, models and setups. We also introduce the research questions, answered by our experiments. 

\subsection{Datasets} 
We use YTSeg \citep{retkowski-waibel-2024-text} as our primary dataset. It contains 19,299 English YouTube videos with their transcripts and chapters. To facilitate more fine-grained analysis of model behavior beyond aggregate metrics, we augment each example with additional annotations. Specifically, every instance is labeled with (i) a \emph{duration regime}, (ii) a \emph{speaker regime}, and (iii) two ASR \emph{transcript variants}. We detail these three annotations in the following and have released them publicly.

\paragraph{Duration Regime.}

We annotate each recording with a duration regime (\Cref{tab:duration-buckets}) for two reasons. First, language model performance has been shown to degrade with increasing context length \cite{liu-etal-2024-lost}. Second, duration reflects computational constraints: many systems cannot process long transcripts. This limitation is particularly acute in audio settings, where audio translates to substantially more model tokens than their text counterparts.

\begin{table}[ht]
  \centering
  \midnotesize
  \begin{tabular}{llr}
    \toprule
    \textbf{Category} & \textbf{Range} & \textbf{Fraction} \\
    \midrule
    Short       & 0--$<$10 min & 38.4\% \\
    Medium      & 10--$<$30 min & 44.2\% \\
    Long        & 30--$<$60 min & 11.1\% \\
    Very long   & $\ge$60 min & \phantom{0}6.3\% \\
    \bottomrule
  \end{tabular}
  \caption{Duration categories and coverage in the dataset.}\vspace{-0.2cm}
  \label{tab:duration-buckets}
\end{table}

\paragraph{Speaker Regime.}

We add a speaker regime to enable analysis by conversational structure, as multi-speaker audio (e.g., speaker changes and overlap) is generally more challenging for speech processing. Videos are categorized as \emph{Single Speaker}, \emph{Weak Single Speaker}, or \emph{Multi Speaker}, see \Cref{tab:speaker-categories} and \Cref{app:speaker_diar} for details on the diarization.

\begin{table}[h]
  \centering
  \midnotesize
  \setlength{\tabcolsep}{2pt}
  \begin{tabular}{llr}
    \toprule
    \textbf{Category} & \textbf{Definition} & \textbf{Fraction} \\
    \midrule
    Single Spk.
      & $N = 1$ \textbf{or} $N \ge 2 \,\wedge\, p_{\text{d}} \ge 0.95$ & 74.2\% \\
    Weak S. Spk. 
      & $N \ge 2 \,\wedge\, 0.8 \le p_{\text{d}} < 0.95$ & 13.0\% \\
    Multi Spk. 
      & $N \ge 2 \,\wedge\, p_{\text{d}} < 0.8$ & 12.9\% \\
    \bottomrule
  \end{tabular}
  \caption{Speaker (Spk.) categories based on number of speakers $N$ and dominant speaker proportion $p_{\text{d}}$.}\vspace{-0.2cm}
  \label{tab:speaker-categories}
\end{table}

\paragraph{Transcripts.}
Finally, we transcribe YTSeg, as the dataset provides only reference transcripts (\Rtag). We use two ASR models with substantially different capacities, Whisper Tiny (\HTtag) and Whisper Large (\HLtag), to study the effect of transcription quality. As expected, ASR\textsubscript{T} yields higher WERs than ASR\textsubscript{L}, see \cref{tab:wer_micro_compare}. This impact also extends to sentence segmentation, where ASR\textsubscript{T} produces coarser segmentation with fewer sentences and segments than Ref and ASR\textsubscript{L} (\Cref{tab:segmentation_comparison}).

\paragraph{Cross-Domain Generalization.} To assess generalization, we additionally evaluate on the AMI corpus \citep{amicorpus}, an out-of-domain benchmark of multi-party meetings whose segmentation differs substantially from YouTube chaptering.

\subsection{Models}

\paragraph{Text-Based Models.} We evaluate text-based segmentation models that operate on transcripts. As our baseline, we use MiniSeg for which we train variants on different transcripts: reference transcripts (\Rtag), ASR transcripts from Whisper Tiny (\HTtag) and Whisper Large (\HLtag), and combinations of them. For zero-shot comparison, we evaluate LLaMA 3.1 8B \cite{grattafiori2024llama3herdmodels} with constrained decoding \cite{retkowski2025paragraph} and WtP\footnote{\huggingfacesmall{} \href{https://huggingface.co/benjamin/wtp-canine-s-12l}{benjamin/wtp-canine-s-12l}} \cite{minixhofer-etal-2023-wheres}.

\paragraph{Text-Based Models with Audio Features.} To assess whether acoustic cues improve text-based segmentation, we augment MiniSeg with hand-crafted audio features. We evaluate: (1) audio features only (without text), (2) text with individual feature categories (pauses, speaking rate, pitch, loudness, speakers), and (3) text with all features combined.

\paragraph{Audio-Only Models.} We develop AudioSeg, a model that operates directly on frame-level acoustic features without requiring transcripts (\Cref{sec:audioseg}). We train and evaluate variants of AudioSeg using different pre-trained audio encoders: HuBERT Base and Large \cite{hsu2021_hubert}, Whisper Large \cite{radford2023}, AF3-Whisper \cite{goel2025audioflamingo3advancing}, Qwen3-AuT \cite{xu2025qwen3omnitechnicalreport}, and PE\textsubscript{A-Frame} Base \cite{vyas2025pushing}.

\paragraph{MLLMs.} For MLLMs, we consider Qwen2.5-Omni\footnote{\huggingfacesmall{} \href{https://huggingface.co/Qwen/Qwen2.5-Omni-7B}{Qwen/Qwen2.5-Omni-7B}}
 \citep{qwen25omni} and Qwen3-Omni\footnote{\huggingfacesmall{} \href{https://huggingface.co/Qwen/Qwen3-Omni-30B-A3B-Instruct}{Qwen/Qwen3-Omni-30B-A3B-Instruct}}
 \citep{xu2025qwen3omnitechnicalreport}, using their default inference parameters. Prompts for all variants are provided in \cref{app:prompts}. For LoRA finetuning, we use Qwen2.5-Omni and the hyperparameters in \Cref{tab:hp_lora}.

\subsection{Evaluation}

\begin{table*}[htb]
\centering
\renewcommand{\arraystretch}{0.75}
\midnotesize
\begin{threeparttable}
\begin{tabular}{lccccccccc}
\toprule
\textbf{Model} &
\multicolumn{3}{c}{\textbf{Evaluated on \Rtag}} &
\multicolumn{3}{c}{\textbf{Evaluated on \HTtag}} &
\multicolumn{3}{c}{\textbf{Evaluated on \HLtag}} \\
\cmidrule(r){2-4} \cmidrule(lr){5-7} \cmidrule(l){8-10}
 & \textbf{F1} (↑) & \textbf{B} (↑) & \textbf{$P_k$} (↓)
 & \textbf{F1} (↑) & \textbf{B} (↑) & \textbf{$P_k$} (↓)
 & \textbf{F1} (↑) & \textbf{B} (↑) & \textbf{$P_k$} (↓) \\
\midrule

\sectionrow{Zero-shot (no fine-tuning)}
\cdashlinelr{1-10}
LLaMA 3.1 8B (Const. Dec.) & 25.92 & 20.69 & 39.98 & 24.71 & 19.91 & 40.60 & 26.26 & 20.78 & 39.75 \\
WtP\textsubscript{canine-s-12l}          & 28.92 & 20.52 & 47.41 & 28.99 & 20.58 & 45.69 & 28.79 & 20.39 & 47.04 \\

\midrule
\sectionrow{Trained on YTSeg (fine-tuned)}
\cdashlinelr{1-10}
MiniSeg \Rtag   & \second{39.54} & 33.21 & 30.13 & 35.87 & 29.15 & 32.09 & 35.58 & 29.39 & 32.14 \\
MiniSeg \HTtag   & 38.40 & 31.32 & 31.20 & 37.30 & \best{30.72} & 31.84 & 36.13 & 29.54 & 32.78 \\
MiniSeg \HLtag   & 37.49 & 30.74 & 31.24 & 35.42 & 28.78 & 32.17 & 35.65 & 29.52 & 31.96 \\

MiniSeg \RHTtag  & \best{40.01} & \best{33.24} & \best{29.84} & \best{37.76} & \second{30.66} & \best{31.29} & \second{36.38} & \second{29.64} & 31.97 \\

MiniSeg \RHLtag  & 39.05 & 32.18 & 30.28 & 35.45 & 28.54 & 32.20 & 35.52 & 29.06 & \second{31.90} \\

MiniSeg \RHTLtag  & \second{39.54} & \second{33.01} & \second{29.96} & \second{37.38} & 30.53 & \second{31.45} & \best{36.52} & \best{30.08} & \best{31.29} \\

\bottomrule
\end{tabular}

% \begin{tablenotes}\footnotesize
% \end{tablenotes}
\end{threeparttable}

\caption{Segmentation performance of text-based models in zero-shot settings and after training on the YTSeg dataset. MiniSeg variants are trained on \Rtag transcripts, \HTtag (Whisper Tiny), \HLtag (Whisper Large), or their combinations, and evaluated on each transcript type.}
\label{tab:asr_seg_results}
\end{table*}

We evaluate all models using Protocol \textbf{T1 (Discrete-time)} with \textit{6-second chunks}, enabling direct comparison across text-based, audio-only, and multimodal models. This chunk duration corresponds to the average sentence length in YTSeg, providing comparable granularity to text-based segmentation. Text-based models and MLLMs are mapped to the time grid via FA of their (predicted) transcripts to extract sentence start timestamps, while AudioSeg operates directly on 6-second windows. We report three metrics computed on the discretized time grid: F1@6s (↑), B@6s (↑), and $P_k$@6s (↓).

\subsection{Experiments}\label{subsec:rq}

We design our experiments to address the following research questions:

\begin{description}[style=unboxed, leftmargin=0pt, parsep=0pt]
    \item[Q1:] \textit{What impact does transcript quality have on text-based segmentation?} We compare segmentation performance using ASR transcripts of varying quality against the reference transcript.

    \item[Q2:] \textit{Does incorporating audio information improve segmentation?} We explore (i)~task-specific models with hand-crafted audio features and learned audio features, and (ii)~MLLMs.

    \item[Q3:] \textit{Beyond segmentation, can MLLMs temporally ground and title chapters?}
    
    \item[Q4:] \textit{How do audio characteristics and domain affect model performance?} We analyze performance as a function of the number of speakers and the duration and validate findings on the AMI dataset.
    
    \item[Q5:] \textit{How reliable and comparable are different evaluation protocols?} We analyze how transcript choice affects metric comparability and validate time-based evaluation.
\end{description}

\section{Results and Analysis}
% This section answers the four research questions introduced in \cref{subsec:rq} and details the results.

\subsection{Q1: What impact does transcript quality have on text-based segmentation?}

\Cref{tab:asr_seg_results} suggests only a weak correspondence between WER and segmentation performance. Zero-shot models (LLaMA 3.1, WtP) that were trained on large corpora demonstrate consistent performance across transcript types, with F1 scores varying by less than 1.5 points. Finetuned MiniSeg models, however, show modest degradation when evaluated on different transcript types than they were trained on. Models trained on reference transcripts lose approximately 3-4 F1 points when evaluated on ASR transcripts. However, joint training on both reference and ASR transcripts substantially improves robustness: MiniSeg \RHTtag achieves the best F1 score on reference transcripts (40.01) while maintaining strong performance on \HTtag (37.76) and \HLtag (36.38). Although Whisper Large achieves lower WER than Whisper Tiny, MiniSeg models trained on \HLtag perform slightly worse than those on \HTtag across most conditions. This indicates that ASR quality does not directly translate to segmentation performance, suggesting other factors beyond word-level accuracy are at play.

\subsection{Q2: Does incorporating audio information improve segmentation?}
% To answer this question, we examine models with audio features, audio-only models, and MLLMs.

\begin{table}[htb]
\centering
\renewcommand{\arraystretch}{0.75}
\midnotesize
\begin{threeparttable}
\begin{tabular}{lccc}
\toprule
\textbf{Variant (MiniSeg \HTtag)} & \textbf{F1} (↑) & \textbf{B} (↑) & \textbf{$P_k$} (↓) \\
\midrule
\multicolumn{4}{l}{\textit{Baselines}} \\
\cdashlinelr{1-4}
Random baseline [\ref{app:random_baseline}]                & \phantom{0}8.57 & \phantom{0}7.90 & 48.43 \\
Audio features only                      & 19.39 & 14.56 & 37.85 \\
Text only                                & 37.30 & 30.72 & 31.84 \\
\midrule
\multicolumn{4}{l}{\textit{Text + single audio feature}} \\
\cdashlinelr{1-4}
+ Speaking Rate \twemoji{stopwatch}        & 37.32 & 30.85 & 31.75 \\
+ Pitch \twemoji{musical note}             & 36.77 & 30.35 & 31.91 \\
+ Loudness \twemoji{speaker high volume}            & 37.82 & 31.02 & 31.50 \\
+ Speakers \twemoji{busts in silhouette}   & 37.97 & 31.11 & 31.48 \\
+ Pauses \twemoji{hourglass}               & \second{40.17} & \best{33.59} & \second{30.25} \\
\midrule
\multicolumn{4}{l}{\textit{Text + all audio features}} \\
\cdashlinelr{1-4}
Feature Combination \twemoji{sparkles}     & \best{40.30} & \second{33.48} & \best{30.35} \\
\bottomrule
\end{tabular}
% \begin{tablenotes}\footnotesize
% \item[1] Random baseline uses the reference number of chapters to generate random segment boundaries.
% \end{tablenotes}
\caption{Segmentation performance of MiniSeg variants trained on \HTtag transcripts.}
\label{tab:audio_features}
\end{threeparttable}
\end{table}

\paragraph{Hand-Crafted Audio Features.} \Cref{tab:audio_features} presents results for MiniSeg trained on \HTtag transcripts augmented with acoustic features. Audio features alone (F1=19.39) substantially outperform the random baseline (F1=8.57), indicating that acoustic cues carry segmentation-relevant information even without semantics. However, the gap to text-only performance (F1=37.30) confirms that semantic content remains essential. When combined with text, specific feature categories yield notable improvements. Pause duration provides the largest gain (+2.87 F1), while speaker features and loudness offer modest benefits, and pitch and speaking rate show no improvement. Combining all features achieves F1=40.30, indicating that the improvements are primarily driven by pause information rather than contributions from multiple features.

\begin{table}[htb]
\centering
\renewcommand{\arraystretch}{0.8}
\midnotesize
\setlength{\tabcolsep}{5pt}
\begin{threeparttable}
\begin{tabular}{p{2cm}ccc}
\toprule
\textbf{Audio Encoder} & \textbf{F1} (↑) & \textbf{B} (↑) & \textbf{$P_k$} (↓) \\
\midrule
Whisper Large   & \best{45.52} & \best{36.17} & \best{28.89} \\
HuBERT Base     & 31.07 & 24.07 & 34.02 \\
HuBERT Large    & 35.58 & 27.95 & 32.23 \\
AF3-Whisper     & \second{39.02} & \second{30.75} & \second{31.23} \\
Qwen3-AuT       & 24.66 & 18.45 & 35.95 \\
PE\textsubscript{A-Frame} Base & 31.36 & 24.33 & 33.27 \\
\bottomrule
\end{tabular}
\caption{Segmentation performance of AudioSeg with different audio encoders.}
\label{tab:audioseg_encoder_comparison}
% \begin{tablenotes}\footnotesize
% \end{tablenotes}
\end{threeparttable}
\end{table}

% Filtered results

% --- (untrained)

% time-chunks-6.00s/accuracy: 94.05 ± 0.14
% time-chunks-6.00s/precision: 49.65 ± 0.91
% time-chunks-6.00s/recall: 38.4 ± 0.8
% time-chunks-6.00s/specificity: 97.71 ± 0.06
% time-chunks-6.00s/f1: 40.27 ± 0.9
% time-chunks-6.00s/pk: 29.6 ± 0.4
% time-chunks-6.00s/window_diff: 31.59 ± 0.42
% time-chunks-6.00s/boundary_similarity: 34.31 ± 0.79
% time-chunks-6.00s/ghd: 1268.3 ± 27.72
% time-chunks-6.00s/num_segments: 5.22 ± 4.16
% reference/num_segments: 8.23 ± 5.76
% reference/num_sentences: 206.32 ± 264.44
% time-chunks-6.00s/f1 [based on p/r]: 43.31 ± 0.62

% --- (trained)

% time-chunks-6.00s/accuracy: 93.97 ± 0.12
% time-chunks-6.00s/precision: 48.67 ± 0.84
% time-chunks-6.00s/recall: 40.12 ± 0.88
% time-chunks-6.00s/specificity: 97.52 ± 0.08
% time-chunks-6.00s/f1: 40.9 ± 0.86
% time-chunks-6.00s/pk: 29.58 ± 0.42
% time-chunks-6.00s/window_diff: 32.16 ± 0.43
% time-chunks-6.00s/boundary_similarity: 34.79 ± 0.76
% time-chunks-6.00s/ghd: 1274.03 ± 29.12
% time-chunks-6.00s/num_segments: 5.56 ± 4.15
% reference/num_segments: 8.23 ± 5.76
% reference/num_sentences: 206.32 ± 264.44
% time-chunks-6.00s/f1 [based on p/r]: 43.98 ± 0.63

\begin{table}[t]
\centering
\renewcommand{\arraystretch}{0.8}
\midnotesize
\setlength{\tabcolsep}{2.9pt}
\renewcommand{\arraystretch}{1.12}
\begin{threeparttable}
\begin{tabular}{@{}l l r r r@{}}
\toprule
\textbf{Model} & \textbf{Strategy\tnote{1}} &
\textbf{F1} (↑) & \textbf{B} (↑) & $\mathbf{P_k}$ (↓) \\
\midrule

\multirow{6}{*}{\textbf{Qwen2.5-Omni}}
  & \DefaultStrat      & 3.43  & 2.34 & 42.67 \\
  & \ICLStrat          & 18.86 & 12.93 & 40.68 \\
  & \ChunkStrat        & 12.14    & 7.40    & 53.51    \\
  & \SelfCasc{\Tonly}  & 9.49    & 6.53    & 42.39    \\
  & \SelfCasc{\TAudio} & 17.96 & 12.52    & 41.22    \\
  & \LoRAStrat               & 24.67    & 17.44    & 38.18    \\

\cdashlinelr{1-5}

\multirow{3}{*}{\textbf{Qwen3-Omni}}
  & \ICLStrat          & \second{41.30}    & \second{35.22}    & \best{33.00}    \\
  & \SelfCasc{\Tonly}  & 21.98    & 18.25    & 36.90    \\
  & \SelfCasc{\TAudio} & 36.14    & 30.51    & \second{34.66}    \\

\midrule

\multicolumn{5}{@{}l}{\textit{Task variant: Segmentation with Timestamps}} \\
\cdashlinelr{1-5}

\multirow{2}{*}{\textbf{Qwen2.5-Omni}}
  & \ICLStrat\,\NoTranscript \textit{(pred)}          & 0.09    & 0.15    & 43.15    \\
  & \LoRAStrat\,\NoTranscript \textit{(pred)}          & 0.21    & 0.45    & 43.40    \\
\cdashlinelr{1-5}
\multirow{3}{*}{\textbf{Qwen3-Omni}}
  & \ICLStrat\,\NoTranscript \textit{(pred)}           & 12.06    & 12.17     & 49.45    \\
  & \ICLStrat\,\WTranscript \textit{(pred)}         & 12.52    & 12.68    & 46.97    \\
  & \ICLStrat\,\WTranscript \textit{(FA)}         & \best{43.84}    & \best{37.83}    & 34.83    \\
\bottomrule
\end{tabular}

\begin{tablenotes}[flushleft]
\footnotesize
\item[1] \Tonly{} = transcript-only, \TAudio{} = transcript+audio.
\item[2] \NoTranscript{} = no transcription, \WTranscript{} = with transcription.
\item[3] \textit{(pred)}: using model-predicted timestamps
\item[4] \textit{(FA)}: obtain timestamps via forced alignment
\end{tablenotes}

\end{threeparttable}
\caption{Segmentation performance for Qwen Omni models, for videos with duration $<$30 minutes.}

\label{tab:qwen-omni-seg-global}
\end{table}

\paragraph{Audio-Only Models.} \Cref{tab:audioseg_encoder_comparison} demonstrates that learned audio representations can achieve competitive performance without any transcript, though encoder choice critically affects results. Whisper Large achieves the best results (F1=45.52), substantially outperforming all text-based models in \Cref{tab:asr_seg_results}. As an ASR encoder, Whisper implicitly captures linguistic structure alongside acoustic patterns, likely explaining its strong performance, possibly aided by domain match, as it was trained on YouTube data similar to YTSeg. HuBERT Large achieves moderate performance (F1=35.58), while PE\textsubscript{A-Frame} (F1=31.36), trained for sound event detection, indicates that non-speech acoustic cues provide some but insufficient signal. Notably, adapting encoders for MLLM pipelines degrades performance: AF3-Whisper (F1=39.02) underperforms Whisper, and Qwen3-AuT (F1=24.66), trained from scratch for Qwen3-Omni without Whisper's foundation, fares worse still. This likely reflects that such encoders are shaped for consumption by an upstream LLM rather than standalone use with lightweight heads, mirroring similar findings in vision~\cite{pmlrzhai,liu2026versavit}. 

\paragraph{MLLMs.} \Cref{tab:qwen-omni-seg-global,tab:omni-failures-wer} report MLLM performance under various inference strategies. Due to context limitations, these results are restricted to videos $<$30 minutes. Qwen2.5-Omni exhibits weak instruction following, frequently dropping transcript segments or entering generation loops. Without adaptation (\DefaultStrat), it achieves only F1=3.43, with 62.41\% of outputs missing transcripts entirely (\Cref{tab:omni-failures-wer}). \ICLStrat improves output quality and segmentation (F1=18.86), eliminating missing transcriptions, but the model still produces only 51.61\% of the expected transcript length. Qwen3-Omni with \ICLStrat demonstrates stronger instruction following (99\% of expected length) and achieves F1=41.30. Comparing \SelfCasc variants reveals the value of audio: Qwen3-Omni chaptering with both transcript and audio (\TAudio: F1=36.14) substantially outperforms text-only (\Tonly: F1=21.98), demonstrating that audio provides complementary signals. \ChunkStrat degrades performance (F1=12.14) due to excessive generation loops, and \LoRAStrat of Qwen2.5-Omni improves segmentation (F1=24.67) but introduces severe hallucinations.

% Total timestamp pairs analyzed: 11621
% Average deviation: 22.04 seconds
% Median deviation: 11.81 seconds
% Min deviation: 0.00 seconds
% Max deviation: 2903.60 seconds

% Deviation distribution:
%   < 5s:  3067 ( 26.4%)
%   < 10s: 5010 ( 43.1%)
%   < 30s: 9303 ( 80.1%)
%   >= 30s: 2318 ( 19.9%)

\paragraph{Non-Speech Cues.} Inspecting the examples where AudioSeg yields the largest gains over text-based models\footnote{E.g., \href{https://www.youtube.com/watch?v=fVFTZu4GaA0}{fVFTZu4GaA0}, \href{https://www.youtube.com/watch?v=unqhMEgBzck}{unqhMEgBzck}, \href{https://www.youtube.com/watch?v=UBzUJB8cLkI}{UBzUJB8cLkI}}, we find a consistent pattern: these videos often contain non-speech audio around chapter transitions, such as music or sound effects. Such cues are only weakly reflected in transcripts, but are directly available to an audio-only model. To confirm this, we filtered out non-speech audio using DeepFilterNet \cite{schroeter2022deepfilternet}, which degraded performance (F1: -2.21, Table~\ref{tab:deepfilternet_ablation}), validating the importance of these acoustic cues.

\subsection{Q3: Beyond segmentation, can MLLMs temporally ground and title chapters?}

\paragraph{Temporal Grounding.} \Cref{tab:qwen-omni-seg-global} also reports results for two task variants that require the model to predict explicit timestamps. Without transcription (\NoTranscript), Qwen3-Omni achieves only F1=12.06, and Qwen2.5-Omni fails almost entirely (F1=0.09). Including transcription (\WTranscript) barely helps when evaluated on predicted timestamps (F1=12.52), revealing that the bottleneck is temporal grounding rather than segmentation ability. This is confirmed by evaluating the same outputs using FA timestamps instead: F1 jumps to 43.84, surpassing even the main \ICLStrat setting (41.30). The predicted timestamps deviate from FA by a median of 11.8\,s (mean: 22.0\,s), with only 43\% falling within 10\,s. These results suggest that while MLLMs can identify topical boundaries, they struggle to localize them precisely in time. Generating a transcript remains essential for reliable temporal grounding via FA.

\paragraph{Title Quality.} \Cref{tab:title-quality} reports title
quality under two protocols (TM, GC; Appendix~\ref{app:title-quality}).
Both protocols capture complementary aspects. For example, Qwen3-Omni \ICLStrat leads on overall quality (GC RL: 26.94) but is comparatively weaker on matched-chapter titles (TM RL: 28.41), whereas Qwen2.5-Omni \LoRAStrat substantially lifts the latter (TM RL: 49.99) without a comparable gain in GC.

\subsection{Q4: How do audio characteristics and domain affect model performance?}
\label{sec:q3}
% We take a closer look at two audio characteristics: the audio duration and the number of speakers.

\begin{figure}[htbp]
    \centering
    \includegraphics[width=0.95\linewidth]{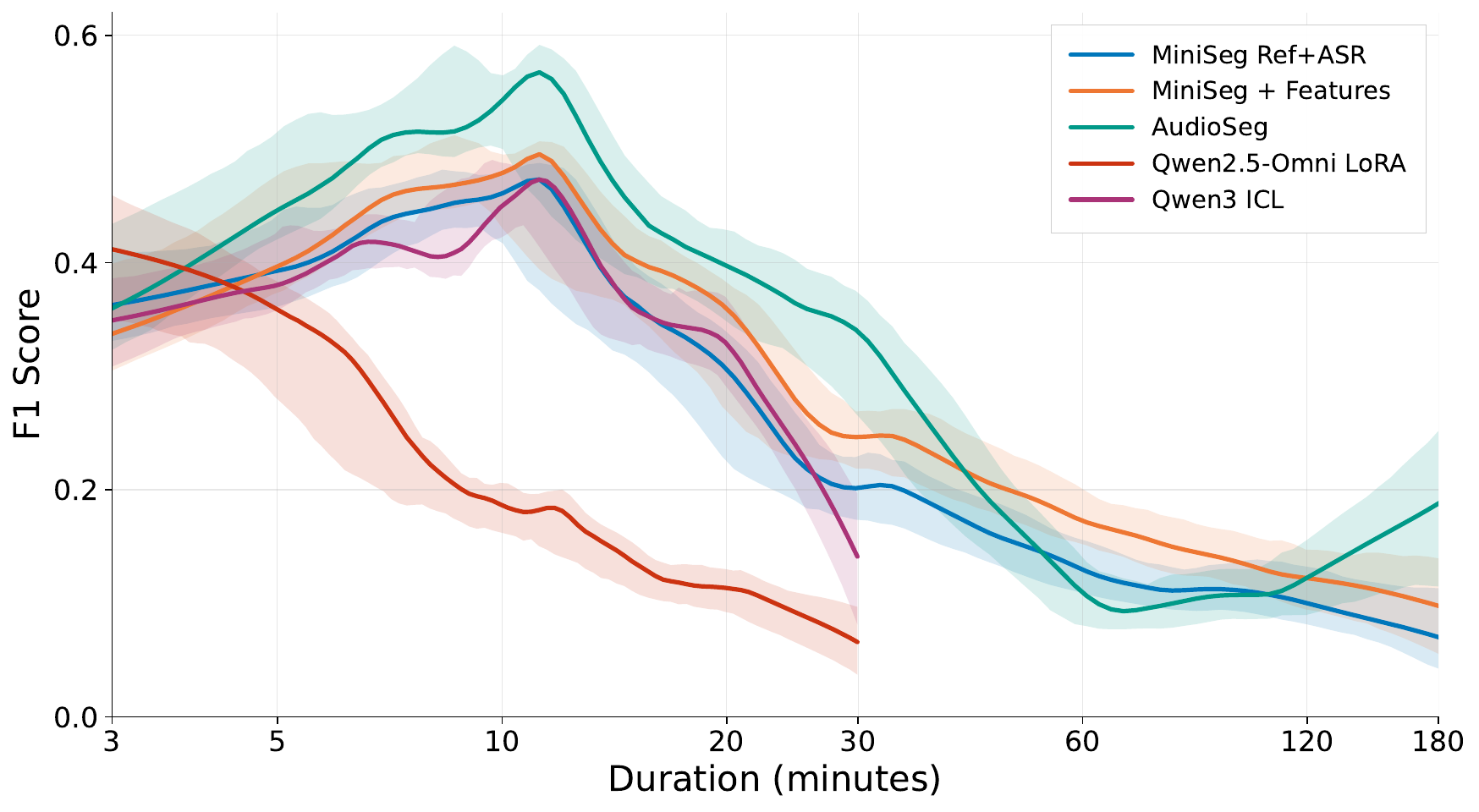}
    \caption{Relation between duration and segmentation performance across models. Smoothed using LOESS.}
    \label{fig:f1_vs_duration_loess}
\end{figure}

\paragraph{Duration Effects.} Figure~\ref{fig:f1_vs_duration_loess} and Table~\ref{tab:seg_by_duration} reveal that AudioSeg clearly dominates for videos $<$30 minutes, peaking at around 10 minutes. Beyond 20--30 minutes, performance degrades sharply for all models while text-based approaches become competitive again: at {$\geq$}60 minutes, MiniSeg + Features (F1=15.17) slightly outperforms AudioSeg (F1=13.28). This convergence at longer durations likely reflects the data distribution: longer videos are underrepresented in training (6.3\% of videos are $\ge$60\,min vs.\ 38.4\% for $<$10\,min), and chapter boundaries become
substantially sparser (0.170 vs.\ 1.178 segments/min; see
\Cref{tab:duration-categories}).

\begin{figure}[htbp]
    \centering
    \includegraphics[width=0.95\linewidth]{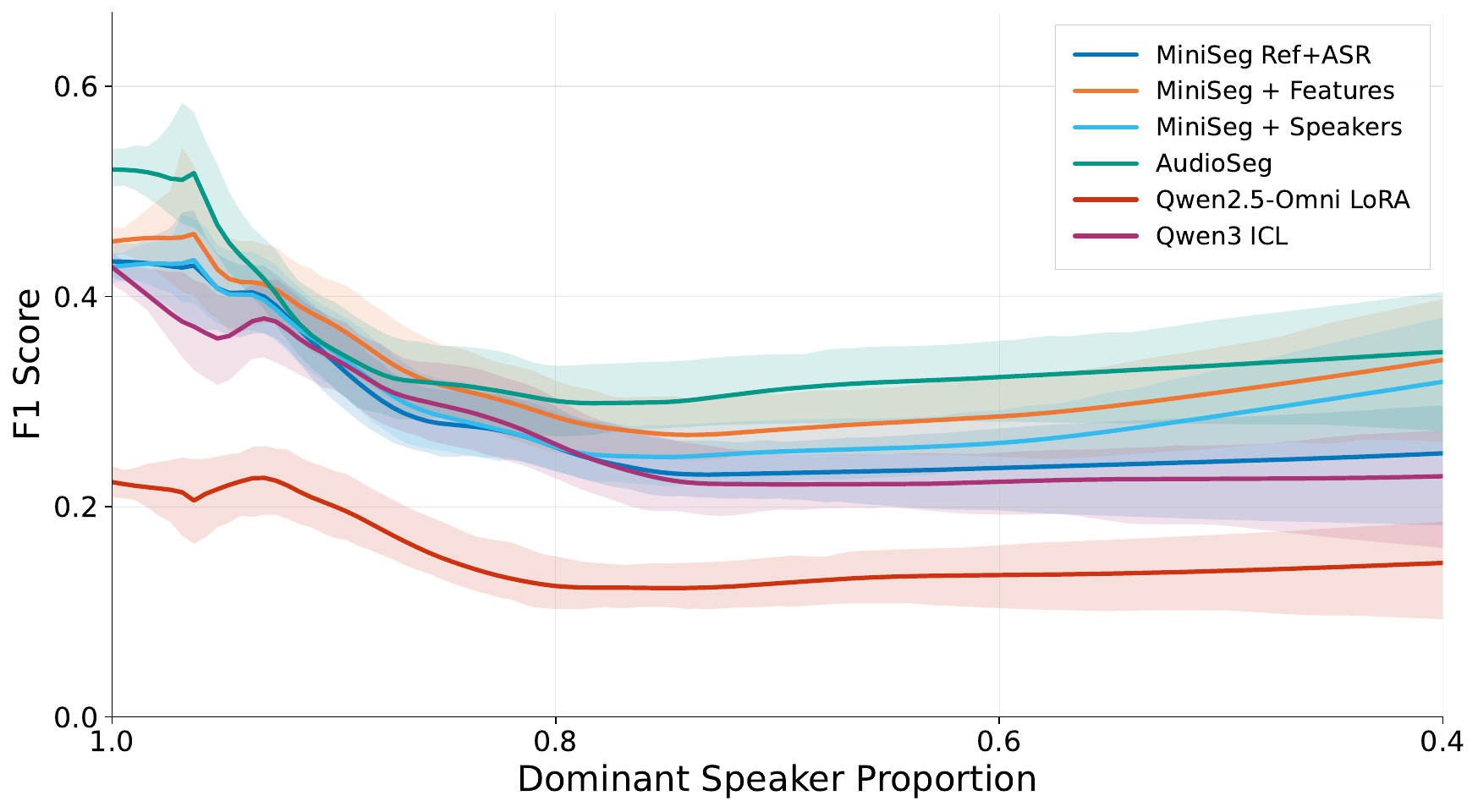}
    \caption{Relation between dominant speaker proportion and segmentation performance across models, for videos $<$30 minutes. Smoothed using LOESS.}
    \label{fig:f1_vs_speaker_loess}
\end{figure}

\paragraph{Speaker Regime Effects.} \Cref{fig:f1_vs_speaker_loess} and \Cref{tab:seg_by_speaker} show that multi-speaker content poses challenges for all approaches, with AudioSeg demonstrating the greatest robustness (F1 dropping from 54.62 to 35.64 between single and multi-speaker settings). Notably, while Table~\ref{tab:audio_features} shows only modest overall gains from speaker features (+0.67 F1), the breakdown by speaker regime reveals their true value: in multi-speaker videos, MiniSeg + Speakers improves F1 from 26.10 to 29.05 (+2.95), whereas single-speaker content is unaffected.

\paragraph{Domain Effects.} \Cref{tab:ami_short,tab:ami_all} show that our main conclusions hold on the AMI dataset: AudioSeg remains the strongest system, and MLLMs do not close the gap. Notably, Qwen3-Omni, which was competitive on YTSeg, performs poorly on AMI, suggesting limited transfer to meeting-style segmentation. Moreover, YTSeg pretraining transfers meaningfully across domains: for both AudioSeg and MiniSeg, the YTSeg→AMI setting consistently outperforms training on AMI alone.

\subsection{Q5: How reliable and comparable are different evaluation protocols?}

\paragraph{Protocol Comparison.} Oracle segmentation (Table~\ref{tab:oracle_segmentation_ceiling}) quantifies the inherent lossiness of transcript-based protocols: even with perfect boundary selection, they only achieve F1≈73–81\%. This discretization loss strongly motivates time-based evaluation protocols that operate directly in continuous time. Protocol comparison (\Cref{tab:protocol_results}) reveals that transcript granularity can inflate text-space metrics for identical predictions due to coarser ASR segmentation. Despite this, Protocol T1 maintains ranking consistency with text-space protocols, validating its use for fair, transcript-invariant comparison across modeling paradigms.

\paragraph{Granularity Sensitivity.} To assess whether our findings depend on the choice of evaluation granularity, we report T1 scores across multiple chunk sizes (6s, 8s, 10s, 12s) in \Cref{tab:t1-chunk-sizes}. Overall, model rankings are consistent and as expected, F1 generally increases with coarser discretization due to wider tolerance windows. We note that AudioSeg exhibits some non-monotonic variation at 8s and 10s chunks; this reflects a re-discretization artifact rather than a true performance change, as the model predicts on a native 6s resolution, and re-binning into non-aligned chunk sizes can cause boundaries to shift across bin edges or collide. To complement T1, we additionally report T2 collar-based F1 (±3s, ±6s) in \Cref{tab:t2-collar}, which operates in continuous time and is therefore free of such discretization artifacts. T2 results show consistent trends and rankings, confirming our conclusions.
\section{Related Work}

Audio chaptering is still an emerging research area that has predominantly been treated as a text segmentation problem operating over transcripts without considering acoustic features \cite{zechner-2002-automatic,Ghazimatin2024,retkowski-waibel-2024-text}. Limited work has explored acoustic features. Earlier work relied on hand-crafted audio features combined with engineered lexical features \cite{lai_automatic_2016,soares2018,lai_integrating_2020}. More recently, \citet{freisinger25_interspeech} studied hierarchical segmentation of long-form videos but incorporated acoustic information only via pauses. At the same time, recent video chaptering work either relies solely on visual cues or emphasizes combining transcripts with visual information, with little attention to audio \cite{yang2023vidchapters,ventura2025}. Separately, the term \emph{audio segmentation} is also used for ASR preprocessing, i.e. partitioning recordings into short, fine-grained chunks, typically via voice activity detection \cite{sohn-1999}, including semantically-informed variants \cite{wetzel-etal-2016-audio,shi23c_interspeech}. Unlike this line of work, we operate on high-level, long-form audio, segmenting for structure into coherent chapters and thereby facing the distinct challenge of modeling long-range context.
\section{Conclusion}
We presented a systematic study of audio chaptering, comparing text-based, audio-only, and multimodal approaches while formalizing evaluation protocols that enable transcript-invariant comparison. Our experiments yield several key findings. First, transcript-free segmentation is not only viable but superior: AudioSeg with Whisper encoders substantially outperforms the best text-based models, demonstrating that learned audio representations capture structural cues beyond what transcripts provide. Second, among acoustic features, pause duration drives nearly all gains when augmenting text-based models, while speaker features provide targeted improvements specifically in multi-speaker content. Third, recent MLLMs such as Qwen3-Omni achieve reasonable ICL performance on shorter audio, though context limitations restrict their applicability. Finally, ASR quality does not directly predict segmentation performance, and joint training on reference and ASR transcripts improves robustness. These findings establish audio chaptering as a task where audio provides complementary or even superior information to text.

\section{Limitations}

Our experiments primarily rely on a single dataset, YTSeg, though we additionally validate our main findings on the small-scale AMI meeting corpus. While YTSeg contains diverse, real-world audio and video content, this reliance may limit the generalizability of our findings. The dataset is also English-only, which restricts conclusions about multilingual applicability. To the best of our knowledge, however, YTSeg is the only publicly available large-scale dataset for audio chaptering, and no comparable datasets exist in other languages that provide similar scale and annotations for this task. Finally, although it would be interesting to fine-tune strong (and bigger) multimodal foundation models such as Qwen3-Omni for this task, we were unable to do so due to computational constraints. Relatedly, while YTSeg naturally extends into the visual modality, we focus on text and audio only; incorporating vision cues (e.g., scene changes, on-screen text, or slide transitions) may further improve performance and remains an important direction for future work.

\section{Potential Risks}
We do not foresee significant risks arising from this work. The primary potential issue is incorrect or imprecise audio chaptering, which may lead to user annoyance, reduced usability, or minor inefficiencies when navigating long audio content. In downstream applications, such errors could cause users to miss relevant sections or expend additional effort locating desired information. However, these consequences are limited in scope and reversible, as chaptering outputs can be corrected or refined without permanent impact. We do not anticipate safety-critical or societal harms resulting from the use of the proposed methods.

\section*{Acknowledgments}
This research is supported by the project ``How is AI Changing Science? Research in the Era of Learning Algorithms'' (HiAICS), funded by the Volkswagen Foundation, and partially by the European Union’s Horizon research and innovation programme under grant agreement No. 101135798, project Meetween (My Personal AI Mediator for Virtual MEETtings BetWEEN People).

\bibliography{custom}

@misc{hu2021loralowrankadaptationlarge,
      title={LoRA: Low-Rank Adaptation of Large Language Models}, 
      author={Edward J. Hu and Yelong Shen and Phillip Wallis and Zeyuan Allen-Zhu and Yuanzhi Li and Shean Wang and Lu Wang and Weizhu Chen},
      year={2021},
      eprint={2106.09685},
      archivePrefix={arXiv},
      primaryClass={cs.CL},
      url={https://arxiv.org/abs/2106.09685}, 
}

@misc{retkowski2025paragraph,
      title={Paragraph Segmentation Revisited: Towards a Standard Task for Structuring Speech}, 
      author={Fabian Retkowski and Alexander Waibel},
      year={2025},
      eprint={2512.24517},
      archivePrefix={arXiv},
      primaryClass={cs.CL},
      url={https://arxiv.org/abs/2512.24517}, 
}

@InProceedings{pmlrzhai,
  title     = {Investigating the Catastrophic Forgetting in Multimodal Large Language Model Fine-Tuning},
  author    = {Zhai, Yuexiang and Tong, Shengbang and Li, Xiao and Cai, Mu and Qu, Qing and Lee, Yong Jae and Ma, Yi},
  booktitle = {Proceedings of the Conference on Parsimony and Learning},
  pages     = {202--227},
  year      = {2024},
  editor    = {Chi, Yuejie and Dziugaite, Gintare Karolina and Qu, Qing and Wang, Atlas Wang and Zhu, Zhihui},
  volume    = {234},
  series    = {Proceedings of Machine Learning Research},
  month     = {03--06 Jan},
  publisher = {PMLR},
  url       = {https://proceedings.mlr.press/v234/zhai24a.html},
}

@article{liu2026versavit,
  title   = {VersaViT: Enhancing MLLM Vision Backbones via Task-Guided Optimization},
  author  = {Liu, Yikun and Liu, Yuan and Di, Shangzhe and Wang, Haicheng and Zhao, Zhongyin and Tian, Le and Zhou, Xiao and Zhou, Jie and Yao, Jiangchao and Wang, Yanfeng and Xie, Weidi},
  journal = {arXiv preprint arXiv:2602.09934},
  year    = {2026}
}

@inproceedings{amicorpus,
title = "The AMI Meeting Corpus: A Pre-announcement",
author = "Jean Carletta and Simone Ashby and Sebastien Bourban and Mike Flynn and Mael Guillemot and Thomas Hain and Jaroslav Kadlec and Vasilis Karaiskos and Wessel Kraaij and Melissa Kronenthal and Guillaume Lathoud and Mike Lincoln and Agnes Lisowska and Iain McCowan and Wilfried Post and Dennis Reidsma and Pierre Wellner",
note = "10.1007/11677482\_3 ; 2nd International Workshop on Machine Learning for Multimodal Interaction, MLMI 2005 ; Conference date: 11-07-2005 Through 13-07-2005",
year = "2006",
doi = "10.1007/11677482\_3",
language = "Undefined",
isbn = "978-3-540-32549-9",
series = "Lecture Notes in Computer Science",
publisher = "Springer",
number = "10",
pages = "28--39",
editor = "S. Renals and S. Bengio",
booktitle = "Machine Learning for Multimodal Interaction, Second International Workshop",
address = "Germany",
}

@inproceedings{zhang2020bertscore,
  title     = {BERTScore: Evaluating Text Generation with {BERT}},
  author    = {Tianyi Zhang and Varsha Kishore and Felix Wu and Kilian Q. Weinberger and Yoav Artzi},
  booktitle = {Proceedings of the International Conference on Learning Representations (ICLR)},
  year      = {2020},
  url       = {https://openreview.net/forum?id=SkeHuCVFDr}
}

@inproceedings{lin-2004-rouge,
    title = "{ROUGE}: A Package for Automatic Evaluation of Summaries",
    author = "Lin, Chin-Yew",
    booktitle = "Text Summarization Branches Out",
    month = jul,
    year = "2004",
    address = "Barcelona, Spain",
    publisher = "Association for Computational Linguistics",
    url = "https://aclanthology.org/W04-1013/",
    pages = "74--81"
}

@inproceedings{zhang2019outline,
author = {Zhang, Ruqing and Guo, Jiafeng and Fan, Yixing and Lan, Yanyan and Cheng, Xueqi},
title = {Outline Generation: Understanding the Inherent Content Structure of Documents},
year = {2019},
isbn = {9781450361729},
publisher = {Association for Computing Machinery},
address = {New York, NY, USA},
url = {https://doi.org/10.1145/3331184.3331208},
doi = {10.1145/3331184.3331208},
booktitle = {Proceedings of the 42nd International ACM SIGIR Conference on Research and Development in Information Retrieval},
pages = {745–754},
numpages = {10},
location = {Paris, France},
series = {SIGIR'19}
}

@inproceedings{wang2020,
  author       = {Wenhui Wang and
                  Furu Wei and
                  Li Dong and
                  Hangbo Bao and
                  Nan Yang and
                  Ming Zhou},
  editor       = {Hugo Larochelle and
                  Marc'Aurelio Ranzato and
                  Raia Hadsell and
                  Maria{-}Florina Balcan and
                  Hsuan{-}Tien Lin},
  title        = {MiniLM: Deep Self-Attention Distillation for Task-Agnostic Compression
                  of Pre-Trained Transformers},
  booktitle    = {Advances in Neural Information Processing Systems 33: Annual Conference
                  on Neural Information Processing Systems 2020, NeurIPS 2020, December
                  6-12, 2020, virtual},
  year         = {2020},
  url          = {https://proceedings.neurips.cc/paper/2020/hash/3f5ee243547dee91fbd053c1c4a845aa-Abstract.html},
  bibsource    = {dblp computer science bibliography, https://dblp.org}
}

@misc{vyas2025pushing,
      title={Pushing the Frontier of Audiovisual Perception with Large-Scale Multimodal Correspondence Learning}, 
      author={Apoorv Vyas and Heng-Jui Chang and Cheng-Fu Yang and Po-Yao Huang and Luya Gao and Julius Richter and Sanyuan Chen and Matt Le and Piotr Dollár and Christoph Feichtenhofer and Ann Lee and Wei-Ning Hsu},
      year={2025},
      eprint={2512.19687},
      archivePrefix={arXiv},
      primaryClass={cs.SD},
      url={https://arxiv.org/abs/2512.19687}, 
}

@inproceedings{minixhofer-etal-2023-wheres,
    title = "Where{'}s the Point? Self-Supervised Multilingual Punctuation-Agnostic Sentence Segmentation",
    author = "Minixhofer, Benjamin  and
      Pfeiffer, Jonas  and
      Vuli{\'c}, Ivan",
    editor = "Rogers, Anna  and
      Boyd-Graber, Jordan  and
      Okazaki, Naoaki",
    booktitle = "Proceedings of the 61st Annual Meeting of the Association for Computational Linguistics (Volume 1: Long Papers)",
    month = jul,
    year = "2023",
    address = "Toronto, Canada",
    publisher = "Association for Computational Linguistics",
    url = "https://aclanthology.org/2023.acl-long.398/",
    doi = "10.18653/v1/2023.acl-long.398",
    pages = "7215--7235"
}

@misc{grattafiori2024llama3herdmodels,
      title={The Llama 3 Herd of Models}, 
      author={Aaron Grattafiori and Abhimanyu Dubey and Abhinav Jauhri and Abhinav Pandey and Abhishek Kadian and Ahmad Al-Dahle and Aiesha Letman and Akhil Mathur and Alan Schelten and Alex Vaughan and Amy Yang and Angela Fan and Anirudh Goyal and Anthony Hartshorn and Aobo Yang and Archi Mitra and Archie Sravankumar and Artem Korenev and Arthur Hinsvark and Arun Rao and Aston Zhang and Aurelien Rodriguez and Austen Gregerson and Ava Spataru and Baptiste Roziere and Bethany Biron and Binh Tang and Bobbie Chern and Charlotte Caucheteux and Chaya Nayak and Chloe Bi and Chris Marra and Chris McConnell and Christian Keller and Christophe Touret and Chunyang Wu and Corinne Wong and Cristian Canton Ferrer and Cyrus Nikolaidis and Damien Allonsius and Daniel Song and Danielle Pintz and Danny Livshits and Danny Wyatt and David Esiobu and Dhruv Choudhary and Dhruv Mahajan and Diego Garcia-Olano and Diego Perino and Dieuwke Hupkes and Egor Lakomkin and Ehab AlBadawy and Elina Lobanova and Emily Dinan and Eric Michael Smith and Filip Radenovic and Francisco Guzmán and Frank Zhang and Gabriel Synnaeve and Gabrielle Lee and Georgia Lewis Anderson and Govind Thattai and Graeme Nail and Gregoire Mialon and Guan Pang and Guillem Cucurell and Hailey Nguyen and Hannah Korevaar and Hu Xu and Hugo Touvron and Iliyan Zarov and Imanol Arrieta Ibarra and Isabel Kloumann and Ishan Misra and Ivan Evtimov and Jack Zhang and Jade Copet and Jaewon Lee and Jan Geffert and Jana Vranes and Jason Park and Jay Mahadeokar and Jeet Shah and Jelmer van der Linde and Jennifer Billock and Jenny Hong and Jenya Lee and Jeremy Fu and Jianfeng Chi and Jianyu Huang and Jiawen Liu and Jie Wang and Jiecao Yu and Joanna Bitton and Joe Spisak and Jongsoo Park and Joseph Rocca and Joshua Johnstun and Joshua Saxe and Junteng Jia and Kalyan Vasuden Alwala and Karthik Prasad and Kartikeya Upasani and Kate Plawiak and Ke Li and Kenneth Heafield and Kevin Stone and Khalid El-Arini and Krithika Iyer and Kshitiz Malik and Kuenley Chiu and Kunal Bhalla and Kushal Lakhotia and Lauren Rantala-Yeary and Laurens van der Maaten and Lawrence Chen and Liang Tan and Liz Jenkins and Louis Martin and Lovish Madaan and Lubo Malo and Lukas Blecher and Lukas Landzaat and Luke de Oliveira and Madeline Muzzi and Mahesh Pasupuleti and Mannat Singh and Manohar Paluri and Marcin Kardas and Maria Tsimpoukelli and Mathew Oldham and Mathieu Rita and Maya Pavlova and Melanie Kambadur and Mike Lewis and Min Si and Mitesh Kumar Singh and Mona Hassan and Naman Goyal and Narjes Torabi and Nikolay Bashlykov and Nikolay Bogoychev and Niladri Chatterji and Ning Zhang and Olivier Duchenne and Onur Çelebi and Patrick Alrassy and Pengchuan Zhang and Pengwei Li and Petar Vasic and Peter Weng and Prajjwal Bhargava and Pratik Dubal and Praveen Krishnan and Punit Singh Koura and Puxin Xu and Qing He and Qingxiao Dong and Ragavan Srinivasan and Raj Ganapathy and Ramon Calderer and Ricardo Silveira Cabral and Robert Stojnic and Roberta Raileanu and Rohan Maheswari and Rohit Girdhar and Rohit Patel and Romain Sauvestre and Ronnie Polidoro and Roshan Sumbaly and Ross Taylor and Ruan Silva and Rui Hou and Rui Wang and Saghar Hosseini and Sahana Chennabasappa and Sanjay Singh and Sean Bell and Seohyun Sonia Kim and Sergey Edunov and Shaoliang Nie and Sharan Narang and Sharath Raparthy and Sheng Shen and Shengye Wan and Shruti Bhosale and Shun Zhang and Simon Vandenhende and Soumya Batra and Spencer Whitman and Sten Sootla and Stephane Collot and Suchin Gururangan and Sydney Borodinsky and Tamar Herman and Tara Fowler and Tarek Sheasha and Thomas Georgiou and Thomas Scialom and Tobias Speckbacher and Todor Mihaylov and Tong Xiao and Ujjwal Karn and Vedanuj Goswami and Vibhor Gupta and Vignesh Ramanathan and Viktor Kerkez and Vincent Gonguet and Virginie Do and Vish Vogeti and Vítor Albiero and Vladan Petrovic and Weiwei Chu and Wenhan Xiong and Wenyin Fu and Whitney Meers and Xavier Martinet and Xiaodong Wang and Xiaofang Wang and Xiaoqing Ellen Tan and Xide Xia and Xinfeng Xie and Xuchao Jia and Xuewei Wang and Yaelle Goldschlag and Yashesh Gaur and Yasmine Babaei and Yi Wen and Yiwen Song and Yuchen Zhang and Yue Li and Yuning Mao and Zacharie Delpierre Coudert and Zheng Yan and Zhengxing Chen and Zoe Papakipos and Aaditya Singh and Aayushi Srivastava and Abha Jain and Adam Kelsey and Adam Shajnfeld and Adithya Gangidi and Adolfo Victoria and Ahuva Goldstand and Ajay Menon and Ajay Sharma and Alex Boesenberg and Alexei Baevski and Allie Feinstein and Amanda Kallet and Amit Sangani and Amos Teo and Anam Yunus and Andrei Lupu and Andres Alvarado and Andrew Caples and Andrew Gu and Andrew Ho and Andrew Poulton and Andrew Ryan and Ankit Ramchandani and Annie Dong and Annie Franco and Anuj Goyal and Aparajita Saraf and Arkabandhu Chowdhury and Ashley Gabriel and Ashwin Bharambe and Assaf Eisenman and Azadeh Yazdan and Beau James and Ben Maurer and Benjamin Leonhardi and Bernie Huang and Beth Loyd and Beto De Paola and Bhargavi Paranjape and Bing Liu and Bo Wu and Boyu Ni and Braden Hancock and Bram Wasti and Brandon Spence and Brani Stojkovic and Brian Gamido and Britt Montalvo and Carl Parker and Carly Burton and Catalina Mejia and Ce Liu and Changhan Wang and Changkyu Kim and Chao Zhou and Chester Hu and Ching-Hsiang Chu and Chris Cai and Chris Tindal and Christoph Feichtenhofer and Cynthia Gao and Damon Civin and Dana Beaty and Daniel Kreymer and Daniel Li and David Adkins and David Xu and Davide Testuggine and Delia David and Devi Parikh and Diana Liskovich and Didem Foss and Dingkang Wang and Duc Le and Dustin Holland and Edward Dowling and Eissa Jamil and Elaine Montgomery and Eleonora Presani and Emily Hahn and Emily Wood and Eric-Tuan Le and Erik Brinkman and Esteban Arcaute and Evan Dunbar and Evan Smothers and Fei Sun and Felix Kreuk and Feng Tian and Filippos Kokkinos and Firat Ozgenel and Francesco Caggioni and Frank Kanayet and Frank Seide and Gabriela Medina Florez and Gabriella Schwarz and Gada Badeer and Georgia Swee and Gil Halpern and Grant Herman and Grigory Sizov and Guangyi and Zhang and Guna Lakshminarayanan and Hakan Inan and Hamid Shojanazeri and Han Zou and Hannah Wang and Hanwen Zha and Haroun Habeeb and Harrison Rudolph and Helen Suk and Henry Aspegren and Hunter Goldman and Hongyuan Zhan and Ibrahim Damlaj and Igor Molybog and Igor Tufanov and Ilias Leontiadis and Irina-Elena Veliche and Itai Gat and Jake Weissman and James Geboski and James Kohli and Janice Lam and Japhet Asher and Jean-Baptiste Gaya and Jeff Marcus and Jeff Tang and Jennifer Chan and Jenny Zhen and Jeremy Reizenstein and Jeremy Teboul and Jessica Zhong and Jian Jin and Jingyi Yang and Joe Cummings and Jon Carvill and Jon Shepard and Jonathan McPhie and Jonathan Torres and Josh Ginsburg and Junjie Wang and Kai Wu and Kam Hou U and Karan Saxena and Kartikay Khandelwal and Katayoun Zand and Kathy Matosich and Kaushik Veeraraghavan and Kelly Michelena and Keqian Li and Kiran Jagadeesh and Kun Huang and Kunal Chawla and Kyle Huang and Lailin Chen and Lakshya Garg and Lavender A and Leandro Silva and Lee Bell and Lei Zhang and Liangpeng Guo and Licheng Yu and Liron Moshkovich and Luca Wehrstedt and Madian Khabsa and Manav Avalani and Manish Bhatt and Martynas Mankus and Matan Hasson and Matthew Lennie and Matthias Reso and Maxim Groshev and Maxim Naumov and Maya Lathi and Meghan Keneally and Miao Liu and Michael L. Seltzer and Michal Valko and Michelle Restrepo and Mihir Patel and Mik Vyatskov and Mikayel Samvelyan and Mike Clark and Mike Macey and Mike Wang and Miquel Jubert Hermoso and Mo Metanat and Mohammad Rastegari and Munish Bansal and Nandhini Santhanam and Natascha Parks and Natasha White and Navyata Bawa and Nayan Singhal and Nick Egebo and Nicolas Usunier and Nikhil Mehta and Nikolay Pavlovich Laptev and Ning Dong and Norman Cheng and Oleg Chernoguz and Olivia Hart and Omkar Salpekar and Ozlem Kalinli and Parkin Kent and Parth Parekh and Paul Saab and Pavan Balaji and Pedro Rittner and Philip Bontrager and Pierre Roux and Piotr Dollar and Polina Zvyagina and Prashant Ratanchandani and Pritish Yuvraj and Qian Liang and Rachad Alao and Rachel Rodriguez and Rafi Ayub and Raghotham Murthy and Raghu Nayani and Rahul Mitra and Rangaprabhu Parthasarathy and Raymond Li and Rebekkah Hogan and Robin Battey and Rocky Wang and Russ Howes and Ruty Rinott and Sachin Mehta and Sachin Siby and Sai Jayesh Bondu and Samyak Datta and Sara Chugh and Sara Hunt and Sargun Dhillon and Sasha Sidorov and Satadru Pan and Saurabh Mahajan and Saurabh Verma and Seiji Yamamoto and Sharadh Ramaswamy and Shaun Lindsay and Shaun Lindsay and Sheng Feng and Shenghao Lin and Shengxin Cindy Zha and Shishir Patil and Shiva Shankar and Shuqiang Zhang and Shuqiang Zhang and Sinong Wang and Sneha Agarwal and Soji Sajuyigbe and Soumith Chintala and Stephanie Max and Stephen Chen and Steve Kehoe and Steve Satterfield and Sudarshan Govindaprasad and Sumit Gupta and Summer Deng and Sungmin Cho and Sunny Virk and Suraj Subramanian and Sy Choudhury and Sydney Goldman and Tal Remez and Tamar Glaser and Tamara Best and Thilo Koehler and Thomas Robinson and Tianhe Li and Tianjun Zhang and Tim Matthews and Timothy Chou and Tzook Shaked and Varun Vontimitta and Victoria Ajayi and Victoria Montanez and Vijai Mohan and Vinay Satish Kumar and Vishal Mangla and Vlad Ionescu and Vlad Poenaru and Vlad Tiberiu Mihailescu and Vladimir Ivanov and Wei Li and Wenchen Wang and Wenwen Jiang and Wes Bouaziz and Will Constable and Xiaocheng Tang and Xiaojian Wu and Xiaolan Wang and Xilun Wu and Xinbo Gao and Yaniv Kleinman and Yanjun Chen and Ye Hu and Ye Jia and Ye Qi and Yenda Li and Yilin Zhang and Ying Zhang and Yossi Adi and Youngjin Nam and Yu and Wang and Yu Zhao and Yuchen Hao and Yundi Qian and Yunlu Li and Yuzi He and Zach Rait and Zachary DeVito and Zef Rosnbrick and Zhaoduo Wen and Zhenyu Yang and Zhiwei Zhao and Zhiyu Ma},
      year={2024},
      eprint={2407.21783},
      archivePrefix={arXiv},
      primaryClass={cs.AI},
      url={https://arxiv.org/abs/2407.21783}, 
}

@inproceedings{brown_fewshot_2020,
 author = {Brown, Tom and Mann, Benjamin and Ryder, Nick and Subbiah, Melanie and Kaplan, Jared D and Dhariwal, Prafulla and Neelakantan, Arvind and Shyam, Pranav and Sastry, Girish and Askell, Amanda and Agarwal, Sandhini and Herbert-Voss, Ariel and Krueger, Gretchen and Henighan, Tom and Child, Rewon and Ramesh, Aditya and Ziegler, Daniel and Wu, Jeffrey and Winter, Clemens and Hesse, Chris and Chen, Mark and Sigler, Eric and Litwin, Mateusz and Gray, Scott and Chess, Benjamin and Clark, Jack and Berner, Christopher and McCandlish, Sam and Radford, Alec and Sutskever, Ilya and Amodei, Dario},
 booktitle = {Advances in Neural Information Processing Systems},
 editor = {H. Larochelle and M. Ranzato and R. Hadsell and M.F. Balcan and H. Lin},
 pages = {1877--1901},
 publisher = {Curran Associates, Inc.},
 title = {Language Models are Few-Shot Learners},
 url = {https://proceedings.neurips.cc/paper_files/paper/2020/file/1457c0d6bfcb4967418bfb8ac142f64a-Paper.pdf},
 volume = {33},
 year = {2020}
}

@inproceedings{titanet22,
  author       = {Nithin Rao Koluguri and
                  Taejin Park and
                  Boris Ginsburg},
  title        = {TitaNet: Neural Model for Speaker Representation with 1D Depth-Wise
                  Separable Convolutions and Global Context},
  booktitle    = {{IEEE} International Conference on Acoustics, Speech and Signal Processing,
                  {ICASSP} 2022, Virtual and Singapore, 23-27 May 2022},
  pages        = {8102--8106},
  publisher    = {{IEEE}},
  year         = {2022},
  url          = {https://doi.org/10.1109/ICASSP43922.2022.9746806},
  doi          = {10.1109/ICASSP43922.2022.9746806},
  bibsource    = {dblp computer science bibliography, https://dblp.org}
}

@article{liu-etal-2024-lost,
    title = "Lost in the Middle: How Language Models Use Long Contexts",
    author = "Liu, Nelson F.  and
      Lin, Kevin  and
      Hewitt, John  and
      Paranjape, Ashwin  and
      Bevilacqua, Michele  and
      Petroni, Fabio  and
      Liang, Percy",
    journal = "Transactions of the Association for Computational Linguistics",
    volume = "12",
    year = "2024",
    address = "Cambridge, MA",
    publisher = "MIT Press",
    url = "https://aclanthology.org/2024.tacl-1.9/",
    doi = "10.1162/tacl_a_00638",
    pages = "157--173"
}

@misc{qwen25omni,
      title={Qwen2.5-Omni Technical Report}, 
      author={Jin Xu and Zhifang Guo and Jinzheng He and Hangrui Hu and Ting He and Shuai Bai and Keqin Chen and Jialin Wang and Yang Fan and Kai Dang and Bin Zhang and Xiong Wang and Yunfei Chu and Junyang Lin},
      year={2025},
      eprint={2503.20215},
      archivePrefix={arXiv},
      primaryClass={cs.CL},
      url={https://arxiv.org/abs/2503.20215}, 
}

@misc{xu2025qwen3omnitechnicalreport,
      title={Qwen3-Omni Technical Report}, 
      author={Jin Xu and Zhifang Guo and Hangrui Hu and Yunfei Chu and Xiong Wang and Jinzheng He and Yuxuan Wang and Xian Shi and Ting He and Xinfa Zhu and Yuanjun Lv and Yongqi Wang and Dake Guo and He Wang and Linhan Ma and Pei Zhang and Xinyu Zhang and Hongkun Hao and Zishan Guo and Baosong Yang and Bin Zhang and Ziyang Ma and Xipin Wei and Shuai Bai and Keqin Chen and Xuejing Liu and Peng Wang and Mingkun Yang and Dayiheng Liu and Xingzhang Ren and Bo Zheng and Rui Men and Fan Zhou and Bowen Yu and Jianxin Yang and Le Yu and Jingren Zhou and Junyang Lin},
      year={2025},
      eprint={2509.17765},
      archivePrefix={arXiv},
      primaryClass={cs.CL},
      url={https://arxiv.org/abs/2509.17765}, 
}

@inproceedings{ventura2025,
  author       = {Lucas Ventura and
                  Antoine Yang and
                  Cordelia Schmid and
                  G{\"{u}}l Varol},
  title        = {Chapter-Llama: Efficient Chaptering in Hour-Long Videos with LLMs},
  booktitle    = {{IEEE/CVF} Conference on Computer Vision and Pattern Recognition,
                  {CVPR} 2025, Nashville, TN, USA, June 11-15, 2025},
  pages        = {18947--18958},
  publisher    = {Computer Vision Foundation / {IEEE}},
  year         = {2025},
  url          = {https://openaccess.thecvf.com/content/CVPR2025/html/Ventura\_Chapter-Llama\_Efficient\_Chaptering\_in\_Hour-Long\_Videos\_with\_LLMs\_CVPR\_2025\_paper.html},
  doi          = {10.1109/CVPR52734.2025.01765},
  bibsource    = {dblp computer science bibliography, https://dblp.org}
}

@article{yurum2023,
author = {Y\"{u}r\"{u}m, Ozan Ra\c{s}it and Ta\c{s}kaya-Temizel, Tu\u{g}ba and Y\i{}ld\i{}r\i{}m, Soner},
title = {The use of video clickstream data to predict university students’ test performance: A comprehensive educational data mining approach},
year = {2022},
issue_date = {May 2023},
publisher = {Kluwer Academic Publishers},
address = {USA},
volume = {28},
number = {5},
issn = {1360-2357},
url = {https://doi.org/10.1007/s10639-022-11403-y},
doi = {10.1007/s10639-022-11403-y},
journal = {Education and Information Technologies},
month = oct,
pages = {5209–5240},
numpages = {32}
}

@article{liao2023,
author = {Liao, Chen-Hsuan and Wu, Jiun-Yu},
title = {Learning analytics on video-viewing engagement in a flipped statistics course: Relating external video-viewing patterns to internal motivational dynamics and performance},
year = {2023},
issue_date = {May 2023},
publisher = {Elsevier Science Ltd.},
address = {GBR},
volume = {197},
number = {C},
issn = {0360-1315},
url = {https://doi.org/10.1016/j.compedu.2023.104754},
doi = {10.1016/j.compedu.2023.104754},
journal = {Comput. Educ.},
month = may,
numpages = {17}
}

@inproceedings{yang2023vidchapters,
  author       = {Antoine Yang and
                  Arsha Nagrani and
                  Ivan Laptev and
                  Josef Sivic and
                  Cordelia Schmid},
  editor       = {Alice Oh and
                  Tristan Naumann and
                  Amir Globerson and
                  Kate Saenko and
                  Moritz Hardt and
                  Sergey Levine},
  title        = {VidChapters-7M: Video Chapters at Scale},
  booktitle    = {Advances in Neural Information Processing Systems 36: Annual Conference
                  on Neural Information Processing Systems 2023, NeurIPS 2023, New Orleans,
                  LA, USA, December 10 - 16, 2023},
  year         = {2023},
  url          = {http://papers.nips.cc/paper\_files/paper/2023/hash/9b5c3e00d6ed30aad7adac9e7a664de1-Abstract-Datasets\_and\_Benchmarks.html},
  bibsource    = {dblp computer science bibliography, https://dblp.org}
}

@inproceedings{soares2018,
author = {Soares, Eduardo R. and Barr\'{e}re, Eduardo},
title = {Automatic Topic Segmentation for Video Lectures Using Low and High-Level Audio Features},
year = {2018},
isbn = {9781450358675},
publisher = {Association for Computing Machinery},
address = {New York, NY, USA},
url = {https://doi.org/10.1145/3243082.3243096},
doi = {10.1145/3243082.3243096},
booktitle = {Proceedings of the 24th Brazilian Symposium on Multimedia and the Web},
pages = {189–196},
numpages = {8},
location = {Salvador, BA, Brazil},
series = {WebMedia '18}
}

@inproceedings{Ghazimatin2024,
  author       = {Azin Ghazimatin and
                  Ekaterina Garmash and
                  Gustavo Penha and
                  Kristen Sheets and
                  Martin Achenbach and
                  Oguz Semerci and
                  Remi Galvez and
                  Marcus Tannenberg and
                  Sahitya Mantravadi and
                  Divya Narayanan and
                  Ofeliya Kalaydzhyan and
                  Douglas Cole and
                  Ben Carterette and
                  Ann Clifton and
                  Paul N. Bennett and
                  Claudia Hauff and
                  Mounia Lalmas},
  editor       = {Edoardo Serra and
                  Francesca Spezzano},
  title        = {{PODTILE:} Facilitating Podcast Episode Browsing with Auto-generated
                  Chapters},
  booktitle    = {Proceedings of the 33rd {ACM} International Conference on Information
                  and Knowledge Management, {CIKM} 2024, Boise, ID, USA, October 21-25,
                  2024},
  pages        = {4487--4495},
  publisher    = {{ACM}},
  year         = {2024},
  url          = {https://doi.org/10.1145/3627673.3680081},
  doi          = {10.1145/3627673.3680081},
  bibsource    = {dblp computer science bibliography, https://dblp.org}
}

@misc{goel2025audioflamingo3advancing,
      title={Audio Flamingo 3: Advancing Audio Intelligence with Fully Open Large Audio Language Models}, 
      author={Arushi Goel and Sreyan Ghosh and Jaehyeon Kim and Sonal Kumar and Zhifeng Kong and Sang-gil Lee and Chao-Han Huck Yang and Ramani Duraiswami and Dinesh Manocha and Rafael Valle and Bryan Catanzaro},
      year={2025},
      eprint={2507.08128},
      archivePrefix={arXiv},
      primaryClass={cs.SD},
      url={https://arxiv.org/abs/2507.08128}, 
}

@article{hsu2021_hubert,
  author       = {Wei{-}Ning Hsu and
                  Benjamin Bolte and
                  Yao{-}Hung Hubert Tsai and
                  Kushal Lakhotia and
                  Ruslan Salakhutdinov and
                  Abdelrahman Mohamed},
  title        = {HuBERT: Self-Supervised Speech Representation Learning by Masked Prediction
                  of Hidden Units},
  journal      = {{IEEE} {ACM} Trans. Audio Speech Lang. Process.},
  volume       = {29},
  pages        = {3451--3460},
  year         = {2021},
  url          = {https://doi.org/10.1109/TASLP.2021.3122291},
  doi          = {10.1109/TASLP.2021.3122291},
  bibsource    = {dblp computer science bibliography, https://dblp.org}
}

@inproceedings{radford2023,
  author       = {Alec Radford and
                  Jong Wook Kim and
                  Tao Xu and
                  Greg Brockman and
                  Christine McLeavey and
                  Ilya Sutskever},
  editor       = {Andreas Krause and
                  Emma Brunskill and
                  Kyunghyun Cho and
                  Barbara Engelhardt and
                  Sivan Sabato and
                  Jonathan Scarlett},
  title        = {Robust Speech Recognition via Large-Scale Weak Supervision},
  booktitle    = {International Conference on Machine Learning, {ICML} 2023, 23-29 July
                  2023, Honolulu, Hawaii, {USA}},
  series       = {Proceedings of Machine Learning Research},
  volume       = {202},
  pages        = {28492--28518},
  publisher    = {{PMLR}},
  year         = {2023},
  url          = {https://proceedings.mlr.press/v202/radford23a.html},
  bibsource    = {dblp computer science bibliography, https://dblp.org}
}

@article{beeferman_statistical_1999,
	title = {Statistical {Models} for {Text} {Segmentation}},
	volume = {34},
	issn = {1573-0565},
	url = {https://doi.org/10.1023/A:1007506220214},
	doi = {10.1023/A:1007506220214},
	language = {en},
	number = {1},
	urldate = {2023-09-19},
	journal = {Machine Learning},
	author = {Beeferman, Doug and Berger, Adam and Lafferty, John},
	month = feb,
	year = {1999},
	pages = {177--210},
}

@inproceedings{retkowski-waibel-2024-text,
    title = "From Text Segmentation to Smart Chaptering: A Novel Benchmark for Structuring Video Transcriptions",
    author = "Retkowski, Fabian  and
      Waibel, Alexander",
    editor = "Graham, Yvette  and
      Purver, Matthew",
    booktitle = "Proceedings of the 18th Conference of the European Chapter of the Association for Computational Linguistics (Volume 1: Long Papers)",
    month = mar,
    year = "2024",
    address = "St. Julian{'}s, Malta",
    publisher = "Association for Computational Linguistics",
    url = "https://aclanthology.org/2024.eacl-long.25/",
    doi = "10.18653/v1/2024.eacl-long.25",
    pages = "406--419"
}

@article{lai_integrating_2020,
	title = {Integrating lexical and prosodic features for automatic paragraph segmentation},
	volume = {121},
	issn = {0167-6393},
	url = {https://www.sciencedirect.com/science/article/pii/S0167639319303541},
	doi = {10.1016/j.specom.2020.04.007},
	urldate = {2025-06-12},
	journal = {Speech Communication},
	author = {Lai, Catherine and Farrús, Mireia and Moore, Johanna D.},
	month = aug,
	year = {2020},
	pages = {44--57},
}

@inproceedings{lai_automatic_2016,
	title = {Automatic {Paragraph} {Segmentation} with {Lexical} and {Prosodic} {Features}},
	doi = {10.21437/Interspeech.2016-992},
	booktitle = {Interspeech 2016},
	author = {Lai, Catherine and Farrús, Mireia and Moore, Johanna D.},
	year = {2016},
	note = {ISSN: 2958-1796},
	pages = {1034--1038},
}

@inproceedings{freisinger25_interspeech,
  title     = {{Towards Multi-Level Transcript Segmentation: LoRA Fine-Tuning for Table-of-Contents Generation}},
  author    = {Steffen Freisinger and Philipp Seeberger and Thomas Ranzenberger and Tobias Bocklet and Korbinian Riedhammer},
  year      = {2025},
  booktitle = {{Interspeech 2025}},
  pages     = {276--280},
  doi       = {10.21437/Interspeech.2025-2792},
  issn      = {2958-1796},
}

@inproceedings{fournier-2013-evaluating,
    title = "Evaluating Text Segmentation using Boundary Edit Distance",
    author = "Fournier, Chris",
    editor = "Schuetze, Hinrich  and
      Fung, Pascale  and
      Poesio, Massimo",
    booktitle = "Proceedings of the 51st Annual Meeting of the Association for Computational Linguistics (Volume 1: Long Papers)",
    month = aug,
    year = "2013",
    address = "Sofia, Bulgaria",
    publisher = "Association for Computational Linguistics",
    url = "https://aclanthology.org/P13-1167/",
    pages = "1702--1712"
}

@article{Li2023,
author = "Xinjian Li",
title = "{Low-Resource Speech Recognition for Thousands of Languages}",
year = "2023",
month = "8",
url = "https://kilthub.cmu.edu/articles/thesis/Low-Resource_Speech_Recognition_for_Thousands_of_Languages/24011307",
doi = "10.1184/R1/24011307.v1"
}

@InProceedings{kurzinger_2020,
author="K{\"u}rzinger, Ludwig
and Winkelbauer, Dominik
and Li, Lujun
and Watzel, Tobias
and Rigoll, Gerhard",
editor="Karpov, Alexey
and Potapova, Rodmonga",
title="CTC-Segmentation of Large Corpora for German End-to-End Speech Recognition",
booktitle="Speech and Computer",
year="2020",
publisher="Springer International Publishing",
address="Cham",
pages="267--278",
isbn="978-3-030-60276-5"
}

@inproceedings{bredin17_interspeech,
  title     = { pyannote.metrics: A Toolkit for Reproducible Evaluation, Diagnostic, and Error Analysis of Speaker Diarization Systems},
  author    = {Hervé Bredin},
  year      = {2017},
  booktitle = {Interspeech 2017},
  pages     = {3587--3591},
  doi       = {10.21437/Interspeech.2017-411},
  issn      = {2958-1796},
}

@Article{2016_mesaros,
AUTHOR = {Mesaros, Annamaria and Heittola, Toni and Virtanen, Tuomas},
TITLE = {Metrics for Polyphonic Sound Event Detection},
JOURNAL = {Applied Sciences},
VOLUME = {6},
YEAR = {2016},
NUMBER = {6},
ARTICLE-NUMBER = {162},
URL = {https://www.mdpi.com/2076-3417/6/6/162},
ISSN = {2076-3417},
DOI = {10.3390/app6060162}
}

@inproceedings{guinaudeau10_interspeech,
  title     = {Improving ASR-based topic segmentation of TV programs with confidence measures and semantic relations},
  author    = {Camille Guinaudeau and Guillaume Gravier and Pascale Sébillot},
  year      = {2010},
  booktitle = {Interspeech 2010},
  pages     = {1365--1368},
  doi       = {10.21437/Interspeech.2010-417},
  issn      = {2958-1796},
}

@inproceedings{schroeter2022deepfilternet,
  title={{DeepFilterNet}: A Low Complexity Speech Enhancement Framework for Full-Band Audio based on Deep Filtering}, 
  author = {Schröter, Hendrik and Escalante-B., Alberto N. and Rosenkranz, Tobias and Maier, Andreas},
  booktitle={ICASSP 2022 IEEE International Conference on Acoustics, Speech and Signal Processing (ICASSP)},
  year={2022},
  organization={IEEE}
}

@inproceedings{lukasik_text_2020,
    address = {Online},
    title = {Text {Segmentation} by {Cross} {Segment} {Attention}},
    url = {https://aclanthology.org/2020.emnlp-main.380},
    doi = {10.18653/v1/2020.emnlp-main.380},
    urldate = {2023-07-24},
    booktitle = {Proceedings of the 2020 {Conference} on {Empirical} {Methods} in {Natural} {Language} {Processing} ({EMNLP})},
    publisher = {Association for Computational Linguistics},
    author = {Lukasik, Michal and Dadachev, Boris and Papineni, Kishore and Simões, Gonçalo},
    month = nov,
    year = {2020},
    pages = {4707--4716},
}

@inproceedings{koshorek_text_2018,
    address = {New Orleans, Louisiana},
    title = {Text {Segmentation} as a {Supervised} {Learning} {Task}},
    url = {https://aclanthology.org/N18-2075},
    doi = {10.18653/v1/N18-2075},
    urldate = {2023-07-24},
    booktitle = {Proceedings of the 2018 {Conference} of the {North} {American} {Chapter} of the {Association} for {Computational} {Linguistics}: {Human} {Language} {Technologies}, {Volume} 2 ({Short} {Papers})},
    publisher = {Association for Computational Linguistics},
    author = {Koshorek, Omri and Cohen, Adir and Mor, Noam and Rotman, Michael and Berant, Jonathan},
    month = jun,
    year = {2018},
    pages = {469--473},
}

@inproceedings{wetzel-etal-2016-audio,
    title = "Audio Segmentation for Robust Real-Time Speech Recognition Based on Neural Networks",
    author = "Wetzel, Micha  and
      Sperber, Matthias  and
      Waibel, Alexander",
    editor = {Cettolo, Mauro  and
      Niehues, Jan  and
      St{\"u}ker, Sebastian  and
      Bentivogli, Luisa  and
      Cattoni, Rolando  and
      Federico, Marcello},
    booktitle = "Proceedings of the 13th International Conference on Spoken Language Translation",
    month = dec # " 8-9",
    year = "2016",
    address = "Seattle, Washington D.C",
    publisher = "International Workshop on Spoken Language Translation",
    url = "https://aclanthology.org/2016.iwslt-1.4/"
}

@ARTICLE{sohn-1999,
  author={Jongseo Sohn and Nam Soo Kim and Wonyong Sung},
  journal={IEEE Signal Processing Letters}, 
  title={A statistical model-based voice activity detection}, 
  year={1999},
  volume={6},
  number={1},
  pages={1-3},
  doi={10.1109/97.736233}}

@inproceedings{shi23c_interspeech,
  title     = {{Semantic VAD: Low-Latency Voice Activity Detection for Speech Interaction}},
  author    = {Mohan Shi and Yuchun Shu and Lingyun Zuo and Qian Chen and Shiliang Zhang and Jie Zhang and Li-Rong Dai},
  year      = {2023},
  booktitle = {{Interspeech 2023}},
  pages     = {5047--5051},
  doi       = {10.21437/Interspeech.2023-598},
  issn      = {2958-1796},
}

@article{zechner-2002-automatic,
    title = "Automatic Summarization of Open-Domain Multiparty Dialogues in Diverse Genres",
    author = "Zechner, Klaus",
    journal = "Computational Linguistics",
    volume = "28",
    number = "4",
    year = "2002",
    address = "Cambridge, MA",
    publisher = "MIT Press",
    url = "https://aclanthology.org/J02-4003/",
    doi = "10.1162/089120102762671945",
    pages = "447--485"
}

@inproceedings{koneru-etal-2026-boom,
    title = "{BOOM}: Beyond Only One Modality {KIT}{'}s Multimodal Multilingual Lecture Companion",
    author = "Koneru, Sai  and
      Retkowski, Fabian  and
      Huber, Christian  and
      Hilgert, Lukas  and
      Akti, Seymanur  and
      Ugan, Enes Yavuz  and
      Waibel, Alexander  and
      Niehues, Jan",
    editor = "Croce, Danilo  and
      Leidner, Jochen  and
      Moosavi, Nafise Sadat",
    booktitle = "Proceedings of the 19th Conference of the {E}uropean Chapter of the {A}ssociation for {C}omputational {L}inguistics (Volume 3: System Demonstrations)",
    month = mar,
    year = "2026",
    address = "Rabat, Marocco",
    publisher = "Association for Computational Linguistics",
    url = "https://aclanthology.org/2026.eacl-demo.14/",
    doi = "10.18653/v1/2026.eacl-demo.14",
    pages = "175--187",
    ISBN = "979-8-89176-382-1"
}

@inproceedings{sarthi-2024,
  author       = {Parth Sarthi and
                  Salman Abdullah and
                  Aditi Tuli and
                  Shubh Khanna and
                  Anna Goldie and
                  Christopher D. Manning},
  title        = {{RAPTOR:} Recursive Abstractive Processing for Tree-Organized Retrieval},
  booktitle    = {The Twelfth International Conference on Learning Representations,
                  {ICLR} 2024, Vienna, Austria, May 7-11, 2024},
  publisher    = {OpenReview.net},
  year         = {2024},
  url          = {https://openreview.net/forum?id=GN921JHCRw},
  bibsource    = {dblp computer science bibliography, https://dblp.org}
}

@inproceedings{retkowski-etal-2025-summarizing,
    title = "Summarizing Speech: A Comprehensive Survey",
    author = {Retkowski, Fabian  and
      Z{\"u}fle, Maike  and
      Sudmann, Andreas  and
      Pfau, Dinah  and
      Watanabe, Shinji  and
      Niehues, Jan  and
      Waibel, Alexander},
    editor = "Christodoulopoulos, Christos  and
      Chakraborty, Tanmoy  and
      Rose, Carolyn  and
      Peng, Violet",
    booktitle = "Proceedings of the 2025 Conference on Empirical Methods in Natural Language Processing",
    month = nov,
    year = "2025",
    address = "Suzhou, China",
    publisher = "Association for Computational Linguistics",
    url = "https://aclanthology.org/2025.emnlp-main.1388/",
    doi = "10.18653/v1/2025.emnlp-main.1388",
    pages = "27275--27306",
    ISBN = "979-8-89176-332-6"
}

\clearpage
\newpage
\appendix
\crefalias{section}{appendix}
\crefalias{subsection}{appendix}
\crefalias{subsubsection}{appendix}

\setcounter{table}{0}
\renewcommand{\thetable}{A\arabic{table}}

\setcounter{figure}{0}
\renewcommand{\thefigure}{A\arabic{figure}}

% ==================== Page 1 ====================
\clearpage
\twocolumn[%
\enlargethispage{3\baselineskip}
\section{Supplementary Results}\label{app:results}

\subsection{Audio Ablation}
\vspace{1em}
\begin{center}
\centering
\renewcommand{\arraystretch}{0.8}
\footnotesize
\setlength{\tabcolsep}{5pt}
\begin{threeparttable}
\captionsetup{width=0.75\linewidth}
\begin{tabular}{p{5cm}ccc}
\toprule
\textbf{Audio Encoder} & \textbf{F1} (↑) & \textbf{B} (↑) & \textbf{$P_k$} (↓) \\
\midrule
Whisper Large   & 45.52 & 36.17 & 28.89 \\
\quad + DeepFilterNet (eval only) & 43.31 & 34.31 & 29.60 \\
\quad + DeepFilterNet (trained) & 43.98 & 34.79 & 29.58 \\
\bottomrule
\end{tabular}
\caption{Performance of AudioSeg when ablated using DeepFilterNet \cite{schroeter2022deepfilternet}, which filters out music and sound effects while preserving human speech.}
\label{tab:deepfilternet_ablation}
\end{threeparttable}
\end{center}

\subsection{Failure Analysis of MLLMs}
\vspace{1em}
\centering
\centering
\small
\setlength{\tabcolsep}{5.5pt}
\renewcommand{\arraystretch}{1.12}
\begin{threeparttable}
\begin{tabular}{@{}l l r r r r r r@{}}
\toprule
\textbf{Model} & \textbf{Strategy\tnote{1}} &
\multicolumn{4}{c}{\textbf{Failure cases (\%)}} &
\multicolumn{2}{c}{\textbf{Transcript (\%)}} \\
\cmidrule(lr){3-6}\cmidrule(lr){7-8}
 &  & \textbf{No Tags} & \textbf{No Transcr.} & \textbf{Empty Chapter\tnote{2}} & \textbf{Gen. Loop\tnote{3}}
 & \textbf{WER} & \textbf{Length} \\
\midrule

\multirow{6}{*}{\textbf{Qwen2.5-Omni}}
  & \DefaultStrat              & 13.12 & 62.41 & 11.03 & 2.17 & 99.98 & 8.86 \\
  & \ICLStrat                  & 15.12 & 0.00  & 1.34  & 2.26 & 93.98 & 51.61 \\
  & \ChunkStrat                & 18.63   & 0.08   & 3.68   & 80.37   & 320.57   & 319.09 \\
  & \SelfCasc{\Tonly}          & 8.60  & 0.25  & 1.67  & 1.67 & 93.02 & 39.05 \\
  & \SelfCasc{\TAudio}         & 5.43  & 0.17  & 1.42  & 2.92 & 93.36 & 56.26 \\
  & \LoRAStrat         & 3.01  & 0.00  & 0.25  & 8.19 & 1032.60 & 1077.88 \\
  
\cdashlinelr{1-8}

\multirow{3}{*}{\textbf{Qwen3-Omni}}
  & \ICLStrat                  & 0.00  & 0.00  & 0.08  & 1.00 & 22.32 & 99.31 \\
  & \SelfCasc{\Tonly}          & 0.08  & 0.00  & 0.58  & 3.26 & 61.92 & 92.62 \\
  & \SelfCasc{\TAudio}         & 0.00  & 0.58  & 0.00  & 0.17 & 31.23 & 87.04 \\
\bottomrule
\end{tabular}

\begin{tablenotes}[flushleft]
\footnotesize
\setlength{\multicolsep}{0pt}
\begin{multicols}{2}
\item[1] \Tonly{} = transcript-only, \TAudio{} = transcript+audio.
\item[2] \textbf{Empty Chapter}: at least one chapter has no transcript.
\item[3] \textbf{Gen. Loop}: a section title/text appears three times.
\end{multicols}
\end{tablenotes}

\end{threeparttable}
\captionof{table}{Failure analysis and transcript-quality metrics for Qwen models under different strategies. We report the rate of four failure modes (missing chapter tags, missing transcription, empty chapters, and generation loops) and transcript quality via WER and relative transcript length. Results are for videos with duration $<$30 minutes.}
\label{tab:omni-failures-wer}

\vspace{1.0em}

\subsection{Stratified Results}
\vspace{1em}
\centering
\centering
\small
\setlength{\tabcolsep}{4.2pt}
\renewcommand{\arraystretch}{1.12}
\begin{threeparttable}
\begin{tabular}{@{}l l r r r@{}}
\toprule
\textbf{Duration} & \textbf{Model} &
\textbf{F1} (↑) & \textbf{B} (↑) & $\mathbf{P_k}$ (↓) \\
\midrule

\multirow{5}{*}{\textbf{$<$10 min}}
  & MiniSeg \RHTtag & 44.06 & 36.98 & 30.00 \\
  & MiniSeg \HTtag\ + Features \twemoji{sparkles} & \second{44.77} & \second{38.22} & \second{29.77} \\
  & AudioSeg (Whisper Large) & \best{50.01} & \best{40.72} & \best{28.08} \\
  & Qwen2.5-Omni \LoRAStrat & 30.33 & 24.13 & 35.47 \\
  & Qwen3-Omni \ICLStrat & 41.80 & 36.12 & 35.49 \\

\cdashlinelr{1-5}

\multirow{5}{*}{\textbf{10--$<$30 min}}
  & MiniSeg \RHTtag & 40.56 & 33.17 & 29.71 \\
  & MiniSeg \HTtag\ + Features \twemoji{sparkles} & \second{44.40} & \second{37.13} & \second{28.22} \\
  & AudioSeg (Whisper Large) & \best{51.40} & \best{41.39} & \best{26.59} \\
  & Qwen2.5-Omni \LoRAStrat & 19.52 & 11.42 & 40.61 \\
  & Qwen3-Omni \ICLStrat & 40.63 & 34.41 & 30.77 \\

\cdashlinelr{1-5}

\multirow{4}{*}{\textbf{30--$<$60 min}}
  & MiniSeg \RHTtag & 17.48 & 11.89 & 37.04 \\
  & MiniSeg \HTtag\ + Features \twemoji{sparkles} & \best{21.89} & \best{15.90} & \second{35.93} \\
  & AudioSeg (Whisper Large) & \second{21.76} & \second{15.28} & \best{35.58} \\
  & Qwen3-Omni \ICLStrat & 12.77 & 9.83 & 39.89 \\

\cdashlinelr{1-5}

\multirow{3}{*}{\textbf{$\geq$60 min}}
  & MiniSeg \RHTtag & 11.76 & 7.19 & 39.98 \\
  & MiniSeg \HTtag\ + Features \twemoji{sparkles} & \best{15.17} & \best{9.79} & \second{38.74} \\
  & AudioSeg (Whisper Large) & \second{13.28} & \second{8.66} & \best{38.10} \\

\bottomrule
\end{tabular}
\end{threeparttable}
\captionof{table}{Segmentation performance across best-performing models by video duration bucket.}
\label{tab:seg_by_duration}
\vspace{0.5em}
]

% ==================== Page 2 ====================
\twocolumn[%
\centering
\centering
\midnotesize
\setlength{\tabcolsep}{4.2pt}
\renewcommand{\arraystretch}{1.12}
\begin{threeparttable}
\begin{tabular}{@{}l l r r r@{}}
\toprule
\textbf{Speaker Category} & \textbf{Model} &
\textbf{F1} (↑) & \textbf{B} (↑) & $\mathbf{P_k}$ (↓) \\
\midrule

\multirow{6}{*}{\textbf{Single Speaker}}
  & MiniSeg \RHTtag & 45.40 & 38.03 & 28.55 \\
  & MiniSeg \HTtag\ + Speakers \twemoji{busts in silhouette} & 44.57 & 37.76 & 29.23 \\
  & MiniSeg \HTtag\ + Features \twemoji{sparkles} & \second{47.19} & \second{40.67} & \second{27.58} \\
  & AudioSeg (Whisper Large) & \best{54.62} & \best{44.99} & \best{25.48} \\
  & Qwen2.5-Omni \LoRAStrat & 26.09 & 18.60 & 38.12 \\
  & Qwen3-Omni \ICLStrat & 44.40 & 38.09 & 31.74 \\

\cdashlinelr{1-5}

\multirow{6}{*}{\textbf{Weak Single Speaker}}
  & MiniSeg \RHTtag & 34.86 & 28.37 & 32.25 \\
  & MiniSeg \HTtag\ + Speakers \twemoji{busts in silhouette} & 35.54 & 29.80 & \second{31.98} \\
  & MiniSeg \HTtag\ + Features \twemoji{sparkles} & \second{38.80} & \best{31.55} & \best{31.50} \\
  & AudioSeg (Whisper Large) & \best{38.99} & 29.40 & 33.12 \\
  & Qwen2.5-Omni \LoRAStrat & 22.89 & 15.60 & 37.72 \\
  & Qwen3-Omni \ICLStrat & 35.55 & \second{30.68} & 34.28 \\

\cdashlinelr{1-5}

\multirow{6}{*}{\textbf{Multi Speaker}}
  & MiniSeg \RHTtag & 26.10 & 19.77 & 36.81 \\
  & MiniSeg \HTtag\ + Speakers \twemoji{busts in silhouette} & 29.05 & 21.16 & \second{35.80} \\
  & MiniSeg \HTtag\ + Features \twemoji{sparkles} & \second{30.89} & \second{22.08} & 36.28 \\
  & AudioSeg (Whisper Large) & \best{35.64} & \best{26.00} & \best{33.76} \\
  & Qwen2.5-Omni \LoRAStrat & 15.34 & 10.71 & 39.27 \\
  & Qwen3-Omni \ICLStrat & 24.53 & 18.70 & 41.20 \\

\bottomrule
\end{tabular}
\end{threeparttable}
\captionof{table}{Segmentation performance by speaker category, for videos with duration $<$30 minutes.}
\label{tab:seg_by_speaker}

\vspace{1.0em}

\subsection{Results on AMI Dataset}

\vspace{1.0em}

\centering
\midnotesize
\setlength{\tabcolsep}{4.2pt}
\renewcommand{\arraystretch}{1.12}
\begin{tabular}{@{}l l r r@{}}
\toprule
\textbf{Model} & \textbf{Pretrain} & \textbf{Val F1} (↑) & \textbf{Test F1} (↑) \\
\midrule
\multirow{2}{*}{AudioSeg}
  & YTSeg $\rightarrow$ AMI & \best{31.19}   & \best{41.91} \\
  & AMI only             & \second{21.12} & \second{21.72} \\
\cdashlinelr{1-4}
\multirow{2}{*}{MiniSeg}
  & YTSeg $\rightarrow$ AMI & 14.68          & 18.41 \\
  & AMI only             & 8.42           & 11.85 \\
\cdashlinelr{1-4}
Qwen2.5-Omni \ICLStrat &   & 9.39  & 7.95 \\
Qwen3-Omni \ICLStrat   &   & 4.99  & 9.16 \\
\bottomrule
\end{tabular}
\captionof{table}{AMI segmentation results for meetings with duration $<$30 minutes (validation: 9, test: 4).}
\label{tab:ami_short}

\vspace{1.0em}

\centering
\midnotesize
\setlength{\tabcolsep}{4.2pt}
\renewcommand{\arraystretch}{1.12}
\begin{tabular}{@{}l l r r@{}}
\toprule
\textbf{Model} & \textbf{Pretrain} & \textbf{Val F1} (↑) & \textbf{Test F1} (↑) \\
\midrule
\multirow{2}{*}{AudioSeg}
  & YTSeg $\rightarrow$ AMI & \best{28.19}   & \best{36.41} \\
  & AMI only             & \second{19.85} & \second{21.96} \\
\cdashlinelr{1-4}
\multirow{2}{*}{MiniSeg}
  & YTSeg $\rightarrow$ AMI & 16.42          & 19.72 \\
  & AMI only             & 7.16           & 13.11 \\
\bottomrule
\end{tabular}
\captionof{table}{AMI segmentation results for all meetings (validation: 16, test: 16).}
\label{tab:ami_all}

% TODO: specify on which transcript was trained/evaluated for MiniSeg
]
\twocolumn[%

\subsection{Evaluation Insights}
\vspace{1em}
\centering

% --- Oracle table (NON-FLOAT) ---
\renewcommand{\arraystretch}{0.8}
\small
\setlength{\tabcolsep}{6pt}
\begin{threeparttable}
\begin{tabular}{llccc}
\toprule
\textbf{Transcript} & \textbf{Timestamps} & \textbf{F1} (↑) & \textbf{B} (↑) & \textbf{$P_k$} (↓) \\
\midrule
\multicolumn{5}{l}{\textit{Native timestamps}} \\
\cdashlinelr{1-5}
\Rtag       & Closed Captions  & \best{80.93} & \best{88.16} & \best{2.35} \\
\HTtag      & Decoder Output   & 79.78 & 85.44 & 3.70 \\
\HLtag      & Decoder Output   & 77.34 & 83.72 & 3.25 \\
\midrule
\multicolumn{5}{l}{\textit{Forced alignment timestamps}} \\
\cdashlinelr{1-5}
\Rtag       & CTC Alignment    & 77.13 & 83.20 & 4.05 \\
\HTtag      & CTC Alignment    & 76.95 & 81.07 & 5.29 \\
\HLtag      & CTC Alignment    & 73.57 & 78.58 & 4.94 \\
\bottomrule
\end{tabular}
\captionof{table}{Oracle segmentation performance ceiling under protocol T1. Oracle places boundaries at the nearest sentence to each true chapter boundary, revealing the inherent discretization loss from projecting continuous-time boundaries onto discrete sentence units.}
\label{tab:oracle_segmentation_ceiling}
\end{threeparttable}

\vspace{1.5em}

\centering
\small
\setlength{\tabcolsep}{4.2pt}
\renewcommand{\arraystretch}{1.12}
\begin{tabular}{@{}l l r r r r@{}}
\toprule
\textbf{Duration} & \textbf{Model / System} &
\textbf{F1@6s} (↑) & \textbf{F1@8s} (↑) & \textbf{F1@10s} (↑) & \textbf{F1@12s} (↑) \\
\midrule
\multirow{6}{*}{\textbf{$<$30 min}}
  & Qwen2.5-Omni \ICLStrat   & 18.86 & 20.52 & 22.08 & 23.28 \\
  & Qwen2.5-Omni \LoRAStrat   & 24.67 & 27.66 & 29.63 & 30.96 \\
  & Qwen3-Omni \ICLStrat      & 41.30 & 44.21 & 46.33 & 48.14 \\
  & AudioSeg (Whisper Large)   & \best{50.75} & \second{46.70} & \second{49.25} & \best{55.61} \\
  & MiniSeg \HTtag + Features \twemoji{sparkles} & \second{44.58} & \best{47.22} & \best{49.34} & \second{51.16} \\
  & MiniSeg \RHTtag      & 42.22 & 44.72 & 46.81 & 48.40 \\
\cdashlinelr{1-6}
\multirow{3}{*}{\textbf{All}}
  & AudioSeg (Whisper Large)   & \best{45.52} & \second{42.01} & \second{44.26} & \best{50.13} \\
  & MiniSeg \HTtag + Features \twemoji{sparkles} & \second{40.30} & \best{42.74} & \best{44.76} & \second{46.49} \\
  & MiniSeg \RHTtag     & 37.76 & 40.09 & 41.97 & 43.42 \\
\bottomrule
\end{tabular}
\captionof{table}{F1 score under the T1 protocol over different time chunk sizes, to assess evaluation granularity sensitivity.}
\label{tab:t1-chunk-sizes}

\vspace{1.5em}

\centering
\small
\setlength{\tabcolsep}{4.2pt}
\renewcommand{\arraystretch}{1.12}
\begin{tabular}{@{}l l r r@{}}
\toprule
\textbf{Duration} & \textbf{Model / System} &
\textbf{F1@$\pm$3s} (↑) & \textbf{F1@$\pm$6s} (↑) \\
\midrule
\multirow{6}{*}{\textbf{$<$30 min}}
  & Qwen2.5-Omni \ICLStrat    & 16.22 & 19.30 \\
  & Qwen2.5-Omni \LoRAStrat   & 20.71 & 24.77 \\
  & Qwen3-Omni \ICLStrat      & 41.12 & 46.65 \\
  & AudioSeg (Whisper Large)  & \best{47.87} & \best{50.90} \\
  & MiniSeg \HTtag + Features \twemoji{sparkles} & \second{43.36} & \second{48.16} \\
  & MiniSeg \RHTtag           & 41.10 & 45.26 \\
\cdashlinelr{1-4}
\multirow{3}{*}{\textbf{All}}
  & AudioSeg (Whisper Large)  & \best{42.50} & \best{46.36} \\
  & MiniSeg \HTtag + Features \twemoji{sparkles} & \second{38.95} & \second{43.50} \\
  & MiniSeg \RHTtag           & 36.42 & 40.27 \\
\bottomrule
\end{tabular}
\captionof{table}{F1 score under the T2 (continuous time) protocol at $\pm$3s and $\pm$6s tolerances.}
\label{tab:t2-collar}

\vspace{1.5em}

% --- Wide results table (NON-FLOAT) ---
\small
\begin{threeparttable}
\begin{tabular}{lcccccc}
\toprule
\textbf{Model} &
\multicolumn{2}{c}{\textbf{Evaluated on \Rtag}} &
\multicolumn{4}{c}{\textbf{Evaluated on \HTtag}} \\
\cmidrule(r){2-3} \cmidrule(l){4-7}
 & \textbf{R1} (↑) & \textbf{T1} (↑)
 & \textbf{H1} (↑) & \textbf{H2} (↑) & \textbf{H3} (↑) & \textbf{T1} (↑) \\
\midrule
MiniSeg \Rtag   & 43.37 & 39.54 & 39.26 & 35.99 & 34.97 & 35.87 \\
MiniSeg \HTtag  & 40.34 & 38.40 & 42.16 & 36.42 & 35.39 & 37.30 \\
MiniSeg \RHTtag & 43.63 & 40.01 & 42.35 & 36.90 & 35.90 & 37.76 \\
\bottomrule
\end{tabular}
\captionof{table}{Protocol comparison for MiniSeg variants (F1 score)}
\label{tab:protocol_results}
\end{threeparttable}
]

\clearpage

\section{Title Quality}
\label{app:title-quality}

% TODO: specify on which transcript was trained/evaluated for BART and LLaMA

\begin{table*}[tbp]
\centering
\small
\setlength{\tabcolsep}{2.9pt}
\renewcommand{\arraystretch}{1.12}
\begin{threeparttable}
\begin{tabular}{@{}l l r r r r r@{}}
\toprule
& & \multicolumn{3}{c}{\textbf{Temporal Matching}} & \multicolumn{2}{c}{\textbf{Global Concat.}} \\
\cmidrule(lr){3-5} \cmidrule(lr){6-7}
\textbf{Model} & \textbf{Strategy} &
\textbf{BS} (↑) & \textbf{RL} (↑) & \textbf{Match} &
\textbf{BS} (↑) & \textbf{RL} (↑) \\
\midrule
\textbf{BART}\tnote{\dag}  & \FullStrat & 92.61 & 48.75 & {--} & 91.47 & 49.95 \\
\textbf{LLaMA 3.1 8B} & \DefaultStrat & 87.70 & 17.38 & 15.86 & 84.42 & 20.77 \\
\cdashlinelr{1-7}
\multirow{2}{*}{\textbf{Qwen2.5-Omni}}
  & \ICLStrat   & \second{91.82} & \second{29.02} & 9.06 & 72.59 & 16.29 \\
  & \LoRAStrat  & \best{92.95}   & \best{49.99} & 12.45 & \second{84.00} & \second{24.74} \\
\textbf{Qwen3-Omni}
  & \ICLStrat   & 89.89 & 28.41 & 28.45 & \best{86.76} & \best{26.94} \\
\bottomrule
\end{tabular}
\begin{tablenotes}[flushleft]
\footnotesize
\item[\dag] From \citet{retkowski-waibel-2024-text}; uses oracle segmentation; trained and evaluated on reference transcripts.
\end{tablenotes}
\end{threeparttable}
\caption{Title quality under the Temporal Matching (TM) and Global Concatenation (GC) protocols, for videos with duration $<$30 minutes. BS = BERTScore F1, RL = ROUGE-L F1, Match = \% of chapters temporally matched.}
\label{tab:title-quality}
\end{table*}

\subsection{Title Evaluation Protocols}

Most work on text segmentation evaluates models solely via boundary quality, even in datasets with section titles \cite{koshorek_text_2018,lukasik_text_2020}. When considered, this is typically done in cascaded pipelines where titles are generated from reference segments, enabling separate evaluation of segmentation and title quality \cite{retkowski-waibel-2024-text}. For end-to-end models that jointly predict boundaries and titles, comparison becomes non-trivial because reference titles are tied to the reference segmentation while predicted titles follow the predicted one. Prior work evaluates titles only when the predicted structure matches the reference \cite{zhang2019outline}. In our setting, boundaries lie in continuous time, so exact-match (EM) is not expected even for qualitatively correct predictions. We therefore propose two protocols that specify how titles are compared. Each protocol induces a set of text pairs to which a text similarity metric $m(\cdot,\cdot)$, such as ROUGE \cite{lin-2004-rouge} or BERTScore \cite{zhang2020bertscore}, can be applied.

\subsubsection{Temporally Matched (TM)}

\textbf{TM} is the temporal analogue of EM. Let $(t_s,t_e,\ell)$ be a reference chapter and $(\hat{t}_s,\hat{t}_e,\hat{\ell})$ the prediction, where $(t_s,t_e)$ is the time interval and $\ell$ the title. A predicted chapter is matched to a reference chapter if boundaries agree within $\delta$ seconds:
\begin{equation}
\left|\hat{t}_{s} - t_{s}\right| \le \delta
\;\;\land\;\;
\left|\hat{t}_{e} - t_{e}\right| \le \delta.
\end{equation}
We compute $m(\hat{\ell}, \ell)$ for each matched pair and average across matched chapters; unmatched chapters are excluded. TM therefore measures title quality on approximately correct temporal structure.

\subsubsection{Global Concatenation (GC)}

\textbf{GC} is structure-independent and evaluates titles at the document level. 
Let $C(\ell_{1:n})$ denote chronological concatenation with newline separators:
\begin{equation}
\label{eq:gc_concat}
C(\ell_{1:n}) \;=\; \ell_{1} \,\Vert\, \texttt{\textbackslash n} \,\Vert\, \cdots \,\Vert\, \texttt{\textbackslash n} \,\Vert\, \ell_{n}.
\end{equation}

We then compute $m\!\big(C(\ell^{p}_{1:L}),\, C(\ell^{g}_{1:K})\big)$. GC is well-defined for $L \neq K$ and captures the collective quality of predicted titles.

\clearpage

\begin{figure*}[!htbp]
\centering
\vspace{-1em}
\section{Prompts}
\label{app:prompts}
\vspace{0.5em}
\centering

\begin{promptpanel}

\prompttitleC{LLaMA 3.1 8B System Prompt (Chaptering)}
\begin{lstlisting}[style=promptcompact]
You are an AI assistant that helps users organize text into high-level chapters with descriptive titles.
\end{lstlisting}

\promptsep

\prompttitleC{LLaMA 3.1 8B User Prompt (Chaptering)}
\begin{lstlisting}[style=promptcompact]
You are tasked with organizing a given text into high-level chapters or sections, each with a descriptive title. The text will be provided to you, and your job is to break it up into a small number of coherent chapters (typically 9 ± 5 chapters per document). Here's the text you'll be working with:

{input}

Your task is to insert chapter breaks with titles into this text. Each chapter should be introduced with a Markdown-style heading in the format: \n\n# Chapter Title\n\n

A chapter is a high-level thematic or functional segment of text that encompasses multiple related ideas or a major narrative/conceptual arc. To identify where to insert chapter breaks and what titles to use, consider the following guidelines:

1. Look for major shifts in topic, theme, or narrative direction
2. Identify distinct phases or stages in the content (e.g., introduction, conclusion)
3. Recognize major transitions between different conceptual areas
4. Consider natural breakpoints that would help a reader understand the overall structure
5. Create descriptive titles that capture the essence of each chapter's content
6. Aim for 9 ± 5 chapters total - fewer, more substantial divisions rather than many small ones

Chapter titles should be:
- Concise but descriptive (2-6 words typically)
- Reflective of the main theme or purpose of that section
- Formatted as proper titles (capitalize important words)

Examples of good chapter titles:
- "Introduction"
- "Historical Background"
- "Main Arguments"
- "Case Study Analysis"
- "Implications"
- "Conclusion"

Please provide your final output with the inserted chapter breaks and titles. Ensure that you maintain the original text exactly as it was given, only adding the chapter headers where appropriate.
\end{lstlisting}

\end{promptpanel}

\caption{LLaMA 3.1 8B system and user prompts for transcript chaptering.}
\label{fig:app_prompts_llama_chap}
\end{figure*}

\begin{figure*}[!htbp]
\centering
\begin{promptpanel}

\prompttitleC{Qwen System Prompt (Chaptering)}
\begin{lstlisting}[style=promptcompact]
You are Qwen, a virtual human developed by the Qwen Team, Alibaba Group, capable of perceiving auditory and visual inputs, as well as generating text and speech. Only return the answer requested. Do not include any explanation or introductions.
\end{lstlisting}

\promptsep

\prompttitleC{Qwen System Prompt (Transcription)}
\begin{lstlisting}[style=promptcompact]
You are Qwen, a virtual human developed by the Qwen Team, Alibaba Group, capable of perceiving auditory and visual inputs, as well as generating text and speech. Only return the answer requested. Do not include any explanation or introductions. You are a speech recognition model.
\end{lstlisting}

\promptsep

\prompttitleC{Qwen User Prompt (Transcription)}
\begin{lstlisting}[style=promptcompact]
Transcribe the English audio into text.
\end{lstlisting}

\end{promptpanel}

\caption{Qwen system prompts and transcription prompt.}
\label{fig:app_prompts_audio_system}
\end{figure*}

\begin{figure*}[!htbp]
\centering
\begin{minipage}[t]{0.49\textwidth}
\prompttitleC{Qwen User Audio Chaptering Prompt}
\begin{promptboxC}
\begin{lstlisting}[style=promptcompact]
You are tasked with organizing a given speech recording into high-level chapters or sections, each with a descriptive title. Your job is to transcribe the speech and break it up into a small number of coherent chapters (typically 9 ± 5 chapters per document).

Each chapter should be introduced in the transcript, but the chapter title must be wrapped in markers for extraction, in the format:

[CSTART] Chapter Title [CEND]

A chapter is a high-level thematic or functional segment of text that encompasses multiple related ideas or a major narrative/conceptual arc. To identify where to insert chapter breaks and what titles to use, consider the following guidelines:

1. Look for major shifts in topic, theme, or narrative direction
2. Identify distinct phases or stages in the content (e.g., introduction, conclusion)
3. Recognize major transitions between different conceptual areas
4. Consider natural breakpoints that would help a reader understand the overall structure
5. Create descriptive titles that capture the essence of each chapter's content
6. Aim for 9 ± 5 chapters total — fewer, more substantial divisions rather than many small ones

Chapter titles should be:
- Concise but descriptive (1-6 words typically)
- Reflective of the main theme or purpose
- Formatted as proper titles (capitalize important words)

Please provide the final transcript with the inserted chapter breaks and titles using the [CSTART] … [CEND] format for all chapter headings.
\end{lstlisting}
\end{promptboxC}
\end{minipage}
\hfill
\begin{minipage}[t]{0.49\textwidth}
\prompttitleC{...with In-context Example}
\begin{promptboxC}
\begin{lstlisting}[style=promptcompact]
You are tasked with organizing a given speech recording into high-level chapters or sections, each with a descriptive title. Your job is to transcribe the speech and break it up into a small number of coherent chapters (typically 9 ± 5 chapters per document).

Each chapter should be introduced in the transcript, but the chapter title must be wrapped in markers for extraction, in the format:

[CSTART] Chapter Title [CEND]

A chapter is a high-level thematic or functional segment of text that encompasses multiple related ideas or a major narrative/conceptual arc. To identify where to insert chapter breaks and what titles to use, consider the following guidelines:

1. Look for major shifts in topic, theme, or narrative direction
2. Identify distinct phases or stages in the content (e.g., introduction, conclusion)
3. Recognize major transitions between different conceptual areas
4. Consider natural breakpoints that would help a reader understand the overall structure
5. Create descriptive titles that capture the essence of each chapter's content
6. Aim for 9 ± 5 chapters total — fewer, more substantial divisions rather than many small ones

Chapter titles should be:
- Concise but descriptive (1-6 words typically)
- Reflective of the main theme or purpose
- Formatted as proper titles (capitalize important words)

Below is an example of how the beginning of a transcript should look with chapter titles inserted using this format:
[CSTART] Introduction [CEND] Welcome everyone, and thank you for joining us today. In this session, we will explore the topic of automatic audio chaptering. [CSTART] Early Background [CEND] Before we dive into the main topic, it's important to understand how the project began...

Please provide the final transcript with the inserted chapter breaks and titles using the [CSTART] … [CEND] format for all chapter headings.
\end{lstlisting}
\end{promptboxC}
\end{minipage}

\caption{Qwen user audio chaptering prompts.}
\label{fig:app_prompts_audio_chap}
\end{figure*}

\begin{figure*}[!htbp]
\centering
\prompttitleC{Qwen User Audio Chaptering Prompt with Chunked Audio Input}
\begin{promptboxC}
\begin{lstlisting}[style=promptcompact]
You are tasked with transcribing an audio chunk and appending it to a given transcript that serves as context. Moreover, the transcribed speech should be broken up into a small number of coherent chapters (typically 9 ± 5 chapters per document).

Each chapter should be introduced in the transcript, but the chapter title must be wrapped in markers for extraction, in the format:

[CSTART] Chapter Title [CEND]

A chapter is a high-level thematic or functional segment of text that encompasses multiple related ideas or a major narrative/conceptual arc. To identify where to insert chapter breaks and what titles to use, consider the following guidelines:

1. Look for major shifts in topic, theme, or narrative direction
2. Identify distinct phases or stages in the content (e.g., introduction, conclusion)
3. Recognize major transitions between different conceptual areas
4. Consider natural breakpoints that would help a reader understand the overall structure
5. Create descriptive titles that capture the essence of each chapter's content
6. Aim for 9 ± 5 chapters total — fewer, more substantial divisions rather than many small ones

Chapter titles should be:
- Concise but descriptive (1-6 words typically)
- Reflective of the main theme or purpose
- Formatted as proper titles (capitalize important words)

Below is an example of how the beginning of a transcript should look with chapter titles inserted using this format:
=== TRANSCRIPT SO FAR (DO NOT REPEAT) ===
[CSTART] Introduction [CEND] Welcome everyone, and thank you for joining us today. This is a very cool educational video and we are delighted that you are watching it.
=== END OF TRANSCRIPT ===
Continue the transcript using ONLY the new audio chunk.\nInsert chapter titles using [CSTART] ... [CEND] as needed.\nNever repeat or restate text from the TRANSCRIPT SO FAR.
Output:
In this session, we will explore the topic of automatic audio chaptering. [CSTART] Early Background [CEND] Before we dive into the main topic, it's important to understand how the project began...

You will receive one audio chunk at the time. Continue the transcript using ONLY the new audio chunk.
=== TRANSCRIPT SO FAR (DO NOT REPEAT) ===
<transcript of previous chunk>
=== END OF TRANSCRIPT ===
Continue the transcript using ONLY the new audio chunk.
Insert chapter titles using [CSTART] ... [CEND] as needed.
Never repeat or restate text from the TRANSCRIPT SO FAR.
\end{lstlisting}
\end{promptboxC}

\caption{Qwen user audio chaptering prompt with chunked input audio (30 s chunks).}
\label{fig:app_prompts_audio_chunking}
\end{figure*}

\begin{figure*}[!htbp]
\centering

\begin{minipage}[t]{0.49\textwidth}
\prompttitleC{Qwen User Audio Chaptering Prompt (Transcript)}
\begin{promptboxC}
\begin{lstlisting}[style=promptcompact]
You are tasked with organizing a given transcript into high-level chapters or sections, each with a descriptive title. Your job is to use the transcription and insert a small number of coherent chapters (typically 9 ± 5 chapters per document).

Each chapter should be introduced in the transcript, but the chapter title must be wrapped in markers for extraction, in the format:

[CSTART] Chapter Title [CEND]

A chapter is a high-level thematic or functional segment of text that encompasses multiple related ideas or a major narrative/conceptual arc. To identify where to insert chapter breaks and what titles to use, consider the following guidelines:

1. Look for major shifts in topic, theme, or narrative direction
2. Identify distinct phases or stages in the content (e.g., introduction, conclusion)
3. Recognize major transitions between different conceptual areas
4. Consider natural breakpoints that would help a reader understand the overall structure
5. Create descriptive titles that capture the essence of each chapter's content
6. Aim for 9 ± 5 chapters total — fewer, more substantial divisions rather than many small ones

Chapter titles should be:
- Concise but descriptive (1-6 words typically)
- Reflective of the main theme or purpose
- Formatted as proper titles (capitalize important words)

Below is an example of how the beginning of a transcript should look with chapter titles inserted using this format:
[CSTART] Introduction [CEND] Welcome everyone, and thank you for joining us today. In this session, we will explore the topic of automatic audio chaptering. [CSTART] Early Background [CEND] Before we dive into the main topic, it's important to understand how the project began...

Please provide the final transcript with the inserted chapter breaks and titles using the [CSTART] … [CEND] format for all chapter headings. Make sure you DO NOT change the transcript itself, only insert chapters.
\end{lstlisting}
\end{promptboxC}
\end{minipage}
\hfill
\begin{minipage}[t]{0.49\textwidth}
\prompttitleC{Qwen User Audio Chaptering Prompt (Transcript + Audio)}
\begin{promptboxC}
\begin{lstlisting}[style=promptcompact]
You are tasked with organizing a given speech recording and corresponding transcript into high-level chapters or sections, each with a descriptive title. Your job is to use the transcription and recording and insert a small number of coherent chapters (typically 9 ± 5 chapters per document).
If the transcript is incorrect at some parts, you can correct it while inserting the chapters.

Each chapter should be introduced in the transcript, but the chapter title must be wrapped in markers for extraction, in the format:

[CSTART] Chapter Title [CEND]

A chapter is a high-level thematic or functional segment of text that encompasses multiple related ideas or a major narrative/conceptual arc. To identify where to insert chapter breaks and what titles to use, consider the following guidelines:

1. Look for major shifts in topic, theme, or narrative direction
2. Identify distinct phases or stages in the content (e.g., introduction, conclusion)
3. Recognize major transitions between different conceptual areas
4. Consider natural breakpoints that would help a reader understand the overall structure
5. Create descriptive titles that capture the essence of each chapter's content
6. Aim for 9 ± 5 chapters total — fewer, more substantial divisions rather than many small ones

Chapter titles should be:
- Concise but descriptive (1-6 words typically)
- Reflective of the main theme or purpose
- Formatted as proper titles (capitalize important words)

Below is an example of how the beginning of a transcript should look with chapter titles inserted using this format:
[CSTART] Introduction [CEND] Welcome everyone, and thank you for joining us today. In this session, we will explore the topic of automatic audio chaptering. [CSTART] Early Background [CEND] Before we dive into the main topic, it's important to understand how the project began...

Please provide the final transcript with the inserted chapter breaks and titles using the [CSTART] … [CEND] format for all chapter headings.
\end{lstlisting}
\end{promptboxC}
\end{minipage}

\caption{Qwen user audio chaptering prompts using transcript-only vs.\ transcript+audio input.}
\label{fig:app_prompts_audio_self_cascaded}
\end{figure*}

\clearpage

\normalsize

\section{Hyperparameters}
\label{app:hyperparameters}

 {
\begin{strip}
\centering
\begin{minipage}[t]{0.48\textwidth}
\vspace{0pt}
\centering
\setcounter{table}{11}
\centering
\begin{tabular}{>{\raggedright\arraybackslash}p{3.75cm}r}
\toprule
\textbf{Hyperparameter} & \textbf{Value} \\
\midrule
\multicolumn{2}{l}{\textit{Architecture}} \\
\cdashlinelr{1-2}
\addlinespace[2pt]
Sentence Encoder & \texttt{\small \huggingfacesmall{} \href{https://huggingface.co/sentence-transformers/all-MiniLM-L12-v2}{all-MiniLM-L12-v2}} \\
Document Encoder & RoFormer \\
\sub Attention Heads & 8 \\
\sub Layers & 12 \\
\sub Embedding Dim & 384 \\
\addlinespace[4pt]
\multicolumn{2}{l}{\textit{Training}} \\
\cdashlinelr{1-2}
\addlinespace[2pt]
Loss Function & Weighted BCE \\
Cross-Entropy Weights & $[1,2]$ \\
Learning Rate & $2.5 \times 10^{-5}$ \\
Batch Size & 115,000 Tokens \\
Epochs & 15 \\
LR Schedule & Cosine \\
Optimizer & AdamW \\
Dropout Rate & 0.1 \\
Gradient Sampling Rate & 0.5 \\
\bottomrule
\end{tabular}
\captionof{table}{Hyperparameters for the architecture and training of MiniSeg}
\label{tab:hyperparameters_miniseg}
\end{minipage}%
\hfill%
\begin{minipage}[t]{0.48\textwidth}
\vspace{0pt}
\centering
\setcounter{table}{12}
\centering
\begin{tabular}{>{\raggedright\arraybackslash}p{4cm}r}
\toprule
\textbf{Hyperparameter} & \textbf{Value} \\
\midrule
\multicolumn{2}{l}{\textit{Architecture}} \\
\cdashlinelr{1-2}
\addlinespace[2pt]
Audio Encoder & \texttt{\small \huggingfacesmall{} \href{https://huggingface.co/openai/whisper-large-v3}{whisper-large-v3}} \\
\sub Chunk Size & 30\,s \\
\sub Batch Chunks & 48 \\
Local Seg. Transformer & Transformer \\
\sub Attention Heads & 4 \\
\sub Layers & 3 \\
\sub Embedding Dim & 384 \\
Document Encoder & RoFormer \\
\sub Attention Heads & 8 \\
\sub Layers & 12 \\
\sub Embedding Dim & 384 \\
\addlinespace[4pt]
\multicolumn{2}{l}{\textit{Training}} \\
\cdashlinelr{1-2}
\addlinespace[2pt]
Loss Function & Weighted BCE \\
Cross-Entropy Weights & $[1,2]$ \\
Learning Rate & $2.5 \times 10^{-5}$ \\
Batch Size & 5\,h \\
Chunk Size & 6\,s \\
Epochs & 5 \\
LR Schedule & Cosine \\
Optimizer & AdamW \\
Dropout Rate & 0.1 \\
Gradient Sampling Rate & 0.5 \\
\bottomrule
\end{tabular}
\captionof{table}{Hyperparameters for the architecture and training of AudioSeg}
\label{tab:hyperparameters_audioseg}
\end{minipage}
\end{strip}
}

\vspace*{-9em}
\setcounter{table}{13}
\centering
\begin{tabular}{>{\raggedright\arraybackslash}p{4.75cm}r}
\toprule
\textbf{Hyperparameter} & \textbf{Value} \\
\midrule
\addlinespace[2pt]
\multicolumn{2}{l}{\textit{LoRA Configuration}} \\
\cdashlinelr{1-2}
\addlinespace[2pt]
Rank & 16 \\
Target Modules & All \\
\addlinespace[4pt]
\multicolumn{2}{l}{\textit{Training}} \\
\cdashlinelr{1-2}
\addlinespace[2pt]
Cutoff Length & 32,768 \\
Batch Size (per device) & 1 \\
Gradient Acc. Steps & 4 \\
Learning Rate & $1.0 \times 10^{-5}$ \\
Epochs & 3 \\
LR Schedule & Cosine \\
Warmup Ratio & 0.1 \\
Precision & FP16 \\
\addlinespace[4pt]
\multicolumn{2}{l}{\textit{Early Stopping}} \\
\cdashlinelr{1-2}
\addlinespace[2pt]
Evaluation Strategy & Steps \\
Patience & 100 Steps \\
Metric & Loss \\
\bottomrule
\end{tabular}
\captionof{table}{Hyperparameters for LoRA \citep{hu2021loralowrankadaptationlarge} finetuning of Qwen 2.5-Omni. We train it for 30 hours on three NVIDIA RTX PRO 6000 Blackwell.}
\label{tab:hp_lora}

\clearpage
\newpage

\justifying

\section{Forced Alignment}

Forced alignment plays a central role in our pipeline, particularly for sentence-level feature extraction. Computing features at the sentence level requires precise start and end timestamps for each sentence. To this end, we employ CTC-based forced alignment using ALQAlign\footnote{\url{https://github.com/xinjli/alqalign}} \cite{kurzinger_2020,Li2023}. Accurate temporal alignment is further essential for our evaluation protocol H3, which assesses time-based overlap between sentences.

\section{Evaluation}

\subsection{Segmentation Evaluation}

We employ the \texttt{segeval}\footnote{\url{https://segeval.readthedocs.io/}} library \cite{fournier-2013-evaluating} to compute segmentation evaluation metrics, including $P_k$ and Boundary Similarity. Although our primary evaluation protocol (T1) operates on time-based chunks rather than individual sentences, these chunks are selected to match the average sentence length. Consequently, we retain the default parameter configurations for both metrics.

\subsection{Alignment Procedures for H2 and H3}
\label{app:alignment}

This appendix details the projection mechanisms used in protocols H2 and H3 to map ASR-based boundary predictions onto the reference transcript $S_{\text{ref}} = (s_1, \dots, s_N)$.

\subsubsection{H2: Word-Level Alignment}

Protocol H2 derives a monotonic mapping from ASR sentences $S_{\text{asr}} = (u_1, \dots, u_M)$ to reference sentences via token-level alignment. For each ASR sentence $u_j$, we identify the reference sentence $s_i$ that contains the majority of its aligned tokens.

A predicted boundary between adjacent ASR sentences $u_j$ and $u_{j+1}$ is projected as follows. If $u_j$ maps to $s_i$ and $u_{j+1}$ maps to $s_{i'}$ with $i' > i$, the boundary is placed at index $i$ in $S_{\text{ref}}$. If both $u_j$ and $u_{j+1}$ map to the same reference sentence $s_i$, a length-based heuristic assigns the boundary to the nearest sentence edge (i.e., $i-1$ or $i$, depending on the relative position of the mapped tokens within $s_i$). This procedure yields a predicted boundary sequence $\hat{\mathbf{y}}$ over $S_{\text{ref}}$, which is then compared to the projected gold boundaries using standard segmentation metrics.

\subsubsection{H3: Time-Based Alignment}

Protocol H3 uses temporal overlap rather than lexical alignment. Given timestamps for both $S_{\text{asr}}$ and $S_{\text{ref}}$ (obtained via ASR decoder output or forced alignment), each ASR sentence $u_j$ is mapped to the reference sentence $s_i$ with which it shares the greatest temporal overlap. In case of zero overlap, $u_j$ is assigned to the reference sentence whose midpoint is nearest to its own. The boundary projection logic then follows H2: boundaries between ASR sentences are transferred to $S_{\text{ref}}$ based on the established index mapping.

\subsection{Random Chaptering Baseline}
\label{app:random_baseline}
This baseline generates random segmentation predictions while preserving the total number of segment boundaries from the reference. For each document, it counts how many boundaries (1s) appear in the reference sequence, then randomly distributes exactly that many boundaries across all possible sentence positions.

\section{Feature Extraction Details}
\label{app:feature_extraction}

This appendix provides a description of the extraction methodology for the hand-crafted audio and speaker features listed in Table~\ref{tab:features}.

\subsection{Features}

\paragraph{Pauses.} The \texttt{pause\_duration} between consecutive sentences was computed as the time gap between the end timestamp of sentence $i-1$ and the start timestamp of sentence $i$. For the first sentence in each video ($i=0$), the pause duration was defined as 0. Negative gaps were clipped to 0 to represent zero pause duration.

\paragraph{Speaking Rate.} Words per minute (WPM) was calculated for each sentence using whitespace tokenization (\texttt{wpm}) by dividing the number of tokens by the sentence duration. To normalize for variability in speaking rates between videos, we additionally computed video-adapted z-scores (\texttt{z\_wpm}).

\paragraph{Pitch.} Fundamental frequency ($F_0$) features were extracted using the YIN algorithm as implemented in \texttt{torchaudio}\footnote{\texttt{torchaudio.functional.detect\_pitch\_frequency}} with a 10~ms frame step, 50--1100~Hz frequency range, and 30-frame median smoothing window. For each sentence with at least 3 voiced frames, we computed absolute features (\texttt{mean\_f0}, \texttt{std\_f0}, \texttt{min\_f0}, \texttt{max\_f0}, \texttt{range\_f0}, \texttt{voicing\_ratio}) and a pitch slope (\texttt{slope\_f0}) via linear regression over time. To account for video-specific pitch ranges, we computed a video-level baseline using the median and standard deviation of $F_0$ across all voiced frames within sentence boundaries, then derived normalized features (\texttt{zF0\_mean}, \texttt{zF0\_slope}). Sentences with fewer than 3 voiced frames were assigned zero values.

\begin{table*}[th]
  \centering
  \small
  \setlength{\tabcolsep}{4pt}
  \begin{tabular}{lrrrrrrr}
    \toprule
    \textbf{Duration Category} & \textbf{Videos \%} & \textbf{Seg. Duration} & \textbf{Seg. Count} & \textbf{Segs/Min} & \textbf{Speaker Count} & \textbf{Dom.\ Spk.\ Prop.} \\
    \midrule
    0--$<$10 min   & 38.4\% & 56.0s (0.9m) & 7.1  & 1.178 & 1.32 & 0.958 \\
    10--$<$30 min  & 44.2\% & 94.6s (1.6m) & 9.7  & 0.681 & 1.46 & 0.933 \\
    30--$<$60 min  & 11.1\% & 282.1s (4.7m) & 9.4  & 0.221 & 1.64 & 0.883 \\
    $\ge$60 min    &  6.3\% & 360.4s (6.0m) & 16.0 & 0.170 & 1.56 & 0.873 \\
    \bottomrule
  \end{tabular}
  \caption{Video duration categories with segment and speaker statistics.}
  \label{tab:duration-categories}
\end{table*}

\paragraph{Loudness.} Loudness features were computed using the ITU-R BS.1770-4 standard as implemented in \texttt{torchaudio}\footnote{\texttt{torchaudio.functional.loudness}}, which measures integrated loudness in LKFS. For each sentence, we extracted the corresponding audio segment and computed its loudness (\texttt{lkfs}). Sentences shorter than 0.4~s were zero-padded before computation. To normalize for video-specific recording levels, we computed video-adapted z-scores (\texttt{z\_loudness}).

\paragraph{Speaker Features.} Speaker diarization was performed using a TitaNet-based pipeline~\cite{titanet22}, which produces timestamped speaker segments. To assign a single speaker label to each sentence, we computed the temporal overlap between the sentence interval and all diarization segments. The speaker with the maximum total overlap duration was assigned to the sentence. In cases where a sentence was fully contained within a single diarization segment, that speaker was assigned directly. For sentences with no overlapping segments, we used the previous sentence's speaker label as a fallback (or the nearest segment by temporal distance for the first sentence). From these per-sentence speaker assignments, we derived seven features: binary flags for speaker continuation (\texttt{same\_as\_prev}) and speaker changes (\texttt{speaker\_change}). The \texttt{turn\_id} feature was computed as the cumulative count of speaker changes. Within-turn features included the 0-indexed position within the current turn (\texttt{pos\_in\_turn}), the total number of sentences in the current turn (\texttt{turn\_len}), and the distance in sentences since the last occurrence of the same speaker (\texttt{dist\_prev\_same}). Finally, \texttt{num\_speakers\_so\_far} tracked the cumulative count of unique speakers encountered from the beginning of the video.

\subsection{Feature Normalization}

We apply global normalization to all features using z-scores computed on the training set:
\begin{equation}
x'_i = \frac{x_i - \mu_{\text{train}}}{\sigma_{\text{train}}}
\end{equation}
where $\mu_{\text{train}}$ and $\sigma_{\text{train}}$ are computed from the training partition and applied to all splits. Features already z-scored within documents (\texttt{z\_wpm}, \texttt{zF0\_*}) are not re-normalized.

\section{Data Annotations}

\subsection{Duration Categories}

\Cref{tab:duration-categories} provides per-category statistics on segment structure and speaker composition. Beyond the training set imbalance and boundary sparsity discussed in \Cref{sec:q3}, longer videos also show more complex speaker structure: the average number of speakers rises from 1.32 to 1.56 and the dominant speaker proportion decreases from 0.958 to 0.873, indicating a shift toward multi-speaker content such as podcasts and interviews.

\subsection{Speaker Diarization}
\label{app:speaker_diar}
We obtain speaker segments using a TitaNet-based diarization pipeline \cite{titanet22}, then define $N$ as the number of speakers with $\ge$10 seconds of total speech and $p_{\text{d}}$ as the dominant speaker's share of total speech time. Based on $(N, p_{\text{d}})$, we are able to categorize the videos into single and multi-speaker settings.

\subsection{Data Statistics for Transcripts}

\vspace{1em}

\begin{center}
\small
\begin{tabular}{lccc}
\toprule
\textbf{Model} & \textbf{Train} & \textbf{Val} & \textbf{Test} \\
\midrule
Whisper Tiny  & 20.17\% & 20.28\% & 19.88\% \\
Whisper Large & 12.81\% & 13.17\% & 12.44\% \\
\bottomrule
\end{tabular}
\captionof{table}{WER of Whisper ASR transcripts on YTSeg}
\label{tab:wer_micro_compare}
\end{center}

\vspace{0.75em}

\begin{center}
\small
\begin{tabular}{lcc}
\toprule
 & \textbf{\# Segments} & \textbf{\# Sentences} \\
\midrule
ASR (Whisper Tiny) & $8.95 \pm 6.45$ & $182.79 \pm 220.62$ \\
ASR (Whisper Large) & $9.13 \pm 6.73$ & $200.48 \pm 282.06$ \\
Reference & $9.18 \pm 6.75$ & $200.23 \pm 287.67$ \\
\bottomrule
\end{tabular}
\captionof{table}{Data statistics on the YTSeg test dataset when applying sentence segmentation on the ASR transcripts versus the reference transcripts.}
\label{tab:segmentation_comparison}
\end{center}

% \subsection{Speech Enhancement}
% We used DeepFilterNet framework \cite{schroeter2022deepfilternet} to do speech enhancement. DeepFilterNet estimates a mask in the time-frequency domain to filter out music, sound effects from the audio and keeps only the human speech. 

\end{document}